\documentclass[acmsmall,authorversion,timestamp,nonacm]{acmart}
\usepackage{subfig}
\usepackage{xcolor,colortbl}
\usepackage[hyphenbreaks]{breakurl}
\usepackage{svg}
\usepackage{pgfplots}
\usetikzlibrary{pgfplots.groupplots}
\usepackage{mathtools}
\usepackage[skip=4pt]{caption}
\usepackage{algorithm}
\usepackage{algpseudocode}

\usepackage[all]{background}
\usepackage{stackengine}
\setstackEOL{\\}
\setstackgap{L}{\normalbaselineskip}
\SetBgContents{\color{gray}{\tiny \Longstack{preprint - accepted by ACM JETC Special Issue \\on Emerging Challenges and Solutions in Hardware Security}}}
\SetBgPosition{3.5cm,1cm}
\SetBgOpacity{1.0}
\SetBgAngle{0}
\SetBgScale{1.8}

\algnewcommand{\LineComment}[1]{\State \textcolor{blue!70}{/* #1 */}}

\acmBooktitle{}

\newcolumntype{L}{>{\centering\arraybackslash}m{0.48\columnwidth}}

\definecolor{ForestGreen}{RGB}{34,139,34} 
\definecolor{Gold}{RGB}{218,165,32} 
\definecolor{MediumBlue}{RGB}{25,25,205} 

\definecolor{bb1}{RGB}{178,24,43}
\definecolor{bb2}{RGB}{244,165,130}
\definecolor{bb3}{RGB}{33,102,172}

\definecolor{c1}{RGB}{171, 171, 171}
\definecolor{c2}{RGB}{217,217,217}
\definecolor{c3}{RGB}{60,179,113}
\usetikzlibrary{patterns}

\definecolor{Cgray}{RGB}{101,104,124}
\definecolor{Cblue}{RGB}{63,136,197}
\definecolor{Cred}{RGB}{244,44,67}
\definecolor{Cgreen}{RGB}{46,160,92}

\definecolor{b0}{RGB}{255,153,153}
\definecolor{b1}{RGB}{235,56,56}
\definecolor{b2}{RGB}{191,0,0}
\definecolor{b2}{RGB}{200,0,0}
\definecolor{b3}{RGB}{100,151,177}
\definecolor{b4}{RGB}{0,91,150}
\definecolor{b5}{RGB}{1,31,75}
\definecolor{gr}{RGB}{102, 102, 102}

\definecolor{orange}{RGB}{250, 154, 80}

\definecolor{darkorange}{RGB}{204,85,0}

\definecolor{col1}{RGB}{5, 45, 135}
\definecolor{col2}{RGB}{14, 112, 209}
\definecolor{col3}{RGB}{57, 180, 232}
\definecolor{col4}{RGB}{221, 40, 9}
\definecolor{col5}{RGB}{240, 139, 0}
\definecolor{col6}{RGB}{118, 192, 0}
\definecolor{col7}{RGB}{116, 134, 148}

\definecolor{dist1}{RGB}{47,79,79}
\definecolor{dist2}{RGB}{139,69,19}
\definecolor{dist3}{RGB}{0,128,0}
\definecolor{dist4}{RGB}{75,0,130}
\definecolor{dist5}{RGB}{255,0,0}
\definecolor{dist6}{RGB}{255,215,0}
\definecolor{dist7}{RGB}{0,255,0}
\definecolor{dist8}{RGB}{0,255,255}
\definecolor{dist9}{RGB}{0,0,255}
\definecolor{dist10}{RGB}{255,0,255}
\definecolor{dist11}{RGB}{100,149,237}
\definecolor{dist12}{RGB}{255,105,180}
\definecolor{dist13}{RGB}{255,140,0}
\DeclarePairedDelimiter\abs{\lvert}{\rvert}

\AtBeginDocument{%
  \providecommand\BibTeX{{%
    \normalfont B\kern-0.5em{\scshape i\kern-0.25em b}\kern-0.8em\TeX}}}

\begin{document}

\title[Challenging the Security of Logic Locking Schemes in the Era of Deep Learning]{Challenging the Security of Logic Locking Schemes in the Era of Deep Learning: A Neuroevolutionary Approach}


\author{Dominik Sisejkovic}
\email{sisejkovic@ice.rwth-aachen.de}

\orcid{0000-0003-3812-727X}
\author{Farhad Merchant}
\email{merchantf@ice.rwth-aachen.de}

\author{Lennart M. Reimann}
\email{reimannl@ice.rwth-aachen.de}

\author{Harshit Srivastava}
\email{srivastava@ice.rwth-aachen.de}

\author{Ahmed Hallawa}
\email{hallawa@ice.rwth-aachen.de}

\author{Rainer Leupers}
\email{leupers@ice.rwth-aachen.de}

\affiliation{%
  \institution{Institute for Communication Technologies and Embedded Systems, RWTH Aachen University, Germany}
  \streetaddress{Templergraben 55}
  \city{Aachen}
  \state{Germany}
  \postcode{52062}
}

\renewcommand{\shortauthors}{D. Sisejkovic et al.}

\begin{abstract}
Logic locking is a prominent technique to protect the integrity of hardware designs throughout the integrated circuit design and fabrication flow. However, in recent years, the security of locking schemes has been thoroughly challenged by the introduction of various deobfuscation attacks. As in most research branches, deep learning is being introduced in the domain of logic locking as well. Therefore, in this paper we present SnapShot: a novel attack on logic locking that is the first of its kind to utilize artificial neural networks to directly predict a key bit value from a locked synthesized gate-level netlist without using a golden reference. Hereby, the attack uses a simpler yet more flexible learning model compared to existing work. Two different approaches are evaluated. The first approach is based on a simple feedforward fully connected neural network. The second approach utilizes genetic algorithms to evolve more complex convolutional neural network architectures specialized for the given task. The attack flow offers a generic and customizable framework for attacking locking schemes using machine learning techniques.
We perform an extensive evaluation of SnapShot for two realistic attack scenarios, comprising both reference combinational and sequential benchmark circuits as well as silicon-proven RISC-V core modules. The evaluation results show that SnapShot achieves an average key prediction accuracy of 82.60\% for the selected attack scenario, with a significant performance increase of 10.49 percentage points compared to the state of the art. Moreover, SnapShot outperforms the existing technique on all evaluated benchmarks. The results indicate that the security foundation of common logic locking schemes is build on questionable assumptions.
Based on the lessons learned, we discuss the vulnerabilities and potentials of logic locking uncovered by SnapShot. The conclusions offer insights into the challenges of designing future logic locking schemes that are resilient to machine learning attacks.
\end{abstract}

\ccsdesc[500]{Security and privacy}
\ccsdesc[500]{Security and privacy~Security in hardware}
\keywords{IP protection, logic locking, deep learning, neuroevolution, RISC-V}

\maketitle

\section{Introduction}
The Integrated Circuit (IC) design and fabrication flow is nowadays heavily steered by short time-to-market and reduced design costs. This trend has led to a horizontal business model in which IC design houses have to rely on third party Intellectual Property (IP) and outsourcing the fabrication to off-site foundries.
The involvement of untrusted parties in both IC design and fabrication has raised countless security concerns, ranging from IP piracy to the insertion of malicious circuit modifications known as hardware Trojans~\cite{trojansLessons,rostami2014primer}.

One promising solution to ensure the integrity of ICs is known as logic locking~\cite{evo2017}. The core idea behind this technique is the insertion of additional logic into a gate-level netlist in order to make the original IP design functionally dependent on a secret key. For example, XOR/XNOR gates can be disseminated in the netlist in a way that the original functionality remains preserved only if a \textit{correct} key drives the additional gates. Since the key is only known to the original IP owner, the IP design remains concealed while passing through the hands of untrusted external (layout) designers and the foundry~\cite{comparativeAnalysis}.
Hereby, the security of logic locking is based on the assumption that a malicious entity has to first uncover the correct activation key before being able to reverse engineer and understand the design. Only thereafter, a meaningful hardware Trojan can be implemented.

In the past decade, a wide spectrum of logic locking schemes~\cite{evo2017, epic2010, sll1} as well as key-recovery attacks~\cite{decadeOfLocking, redundancyIdentification2019, SAIL2019, SURF2019, BOCANet2019, sat2015} have been introduced. However, the efficiency and applicability of these attacks greatly depend on the assumed attack model. Most attacks in the past have relied on the availability of a golden reference, i.e., an \textit{oracle} with I/O access in the form of an activated IC (available from the semiconductor market) or a set of golden I/O patterns (e.g., test vectors). Based on this assumption, powerful oracle-guided attacks have been devised, such as Boolean Satisfiability (SAT) based attacks~\cite{decadeOfLocking} and path-sensitization~\cite{sll1}.

\textbf{Motivation:}~Several important observations indicate the diminishing applicability of oracle-guided attacks. First, it has recently been shown that having physical access to an activated IC allows the adversary to get hold of the secret key, even in the presence of a tamper-proof memory~\cite{KeyReading2019}. Secondly, it is possible to design hardware Trojans that leak the secret key to an adversary once the IC is activated, regardless of the locking scheme or key protection mechanism~\cite{TAAL2019}. These two observations originate in the problem of storing and protecting the secret logic locking key on the fabricated chip. Finally, the most powerful attacks, including SAT-based attacks, often rely on having full access to an activated IC, including an open scan-chain; otherwise, most attacks become infeasible~\cite{decadeOfLocking}. Interestingly, this assumption is not realistic, since genuine IC vendors never leave a scan-chain open or at least use some form of authentication~\cite{secureScanChain}. These observations yield two important conclusions: ($i$) currently, the only realistic attacks on logic locking are oracle-less attacks, since the oracle offers no advantage due to the mentioned pitfalls and ($ii$) the primary focus of logic locking is shifting towards the protection of an IP design in its first production round before an oracle is available. Despite the inherent difficulty of not having a golden reference, a powerful oracle-less attack known as SAIL has recently been reported~\cite{SAIL2019}. SAIL aims at recovering local logic structures from a locked IC using Machine Learning (ML) without the necessity of oracle I/O access. The success of SAIL emerges out of the simple, predictable and often deterministic changes introduced by logic locking schemes. 
With the introduction of SAIL, a new ground for oracle-less attacks on logic locking using ML techniques has been established. 

\textbf{Contributions:} Inspired by SAIL, this work introduces \textit{SnapShot}; a novel oracle-less attack on logic locking that utilizes Artificial Neural Networks (ANNs) to directly predict key values by analyzing "snapshots", i.e., excerpts of a locked gate-level netlist. The significance of SnapShot lies in the following:
\begin{itemize}
	\item SnapShot is applicable without a golden reference, thereby following a realistic attack scenario. The adversary only needs access to the gate-level netlist or layout, upon which the netlist can be extracted. 
	\item The attack is applicable for any known gate-insertion-based scheme. The general applicability of the SnapShot attack flow is provided by the representation used for processing netlists with a machine learning model. This representation is capable of incorporating any kind of netlist structures through customization (elaborated in Section~\ref{snapshot:extraction}).
	\item SnapShot is suitable for both combinational and sequential ICs. This is another feature provided by the netlist representation format which is easily expandable to incorporate flip-flops in sequential ICs (evaluated in Section~\ref{results}).
	\item SnapShot has a linear time complexity with respect to the key length. This feature is given by the nature of the attack: each key input can be processed separately.	
	\item The attack flow can be customized for attack exploration: any supervised learning model can be used for the training process irrespective of the locking scheme (Section~\ref{snapshot:evolution}).
\end{itemize}
Based on the aforementioned advantages, the key contributions of this work can be summarized as follows:
\begin{itemize}
	\item The introduction of SnapShot: an oracle-less attack on logic locking that is the first of its kind to \textit{directly} predict a key value from a locked synthesized gate-level netlist thereby utilizing neuroevolutionary and deep learning techniques. 
	The applicability of SnapShot is showcased using two approaches. The first one is based on a simple neural network with fully connected layers. The second approach utilizes a neuroevolutionary mechanism to automatically evolve Convolutional Neural Network (CNN) architectures with genetic algorithms, thereby mitigating the complexity of manually drafting suitable CNN architectures. The evolved networks are targeted towards maximizing the attack efficiency for specific benchmarks and attack scenarios. SnapShot achieves an average activation key prediction accuracy of 82.60\%, thereby outperforming the state-of-the-art attack by 10.49 percentage points.
	\item A new representation of a selected subcircuit that is expandable to sequential ICs and adaptive in terms of how much information is stored for a given circuit. This representation allows the processing of any gate-insertion-based logic locking scheme and is well suited for the training of ANNs.
	\item The evaluation of  SnapShot on two realistic attack scenarios, including both commonly used benchmarks and silicon-proven RISC-V processor modules. To the best of our knowledge, the attack scenario based on a generalized set of netlists has not been evaluated before.
	\item An analysis of the emerging challenges in the design of logic locking schemes that are resilient to ML-based key-guessing attacks.
\end{itemize}
The rest of the paper is organized as follows. Section~\ref{preliminaries} introduces the basic theoretical concepts used throughout this work. Section~\ref{related} presents the related work. The SnapShot attack flow is presented in Section~\ref{snapshot}, and the experimental evaluation and results are discussed in Section~\ref{results}. A discussion on the challenges of ML-resilient locking is given in Section~\ref{discussion}. Limitations and research opportunities are outlined in Section~\ref{futurework}. Finally, Section~\ref{conclusion} concludes the paper. 

\section{Preliminaries} \label{preliminaries}
\subsection{Logic Locking}\label{intro:locking}
Logic locking is used for protecting the integrity of ICs throughout the design and fabrication flow. 
Usually, logic locking is implemented by inserting different types of key-bounded gates (known as \textit{key gates}) into a netlist. The key bit for each key gate is provided through the input pins of the netlist. For example, the random insertion scheme known as EPIC disseminates XOR/XNOR gates at random locations in the netlist~\cite{epic2010}. XOR gates bound to a key bit 0 and XNOR gates bound to a key bit 1 always buffer the second gate input, thereby preserving the original IC functionality. One example is given in Fig.~\ref{fig:LL}~(a). Here the key gate KG1 is added into the netlist. If driven by the correct key bit ($k_{1}=0$), the signal $s$ is buffered through KG1, thereby preserving its original value. The \textit{security foundation} of this scheme lies in the difficulty of guessing whether an inverter of the XNOR or XOR+INV gate belongs to the original circuit or not; otherwise, a simple removal would be possible.
In the past decade, a wide range of locking schemes has been introduced~\cite{evo2017}, all relying on the insertion of some form of key-controlled redundant hardware into a netlist.

\subsection{Logic Locking in the IC Design and Fabrication Flow} \label{preliminaries:llinicflow}
\begin{figure}[!t]
	\centering
	\subfloat[]{
		\includegraphics[width=0.3\columnwidth]{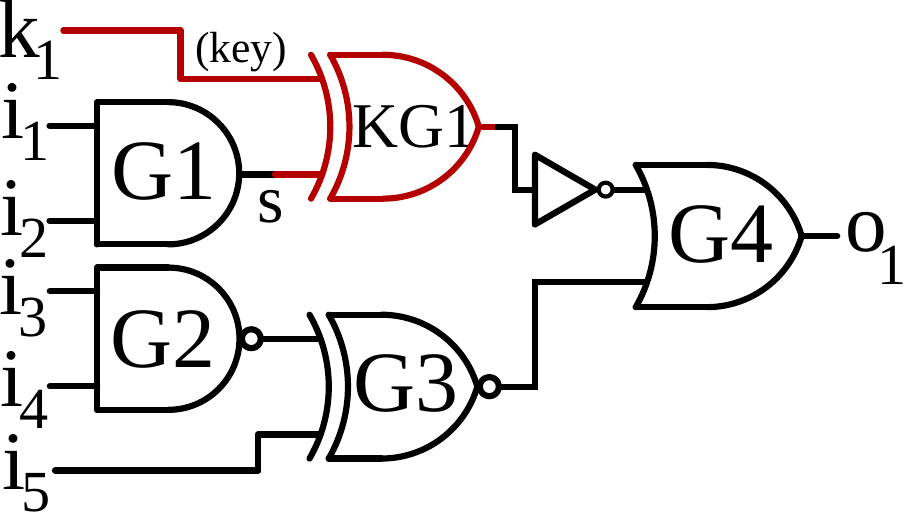}
	}
	\subfloat[]{
		\includegraphics[width=0.68\columnwidth]{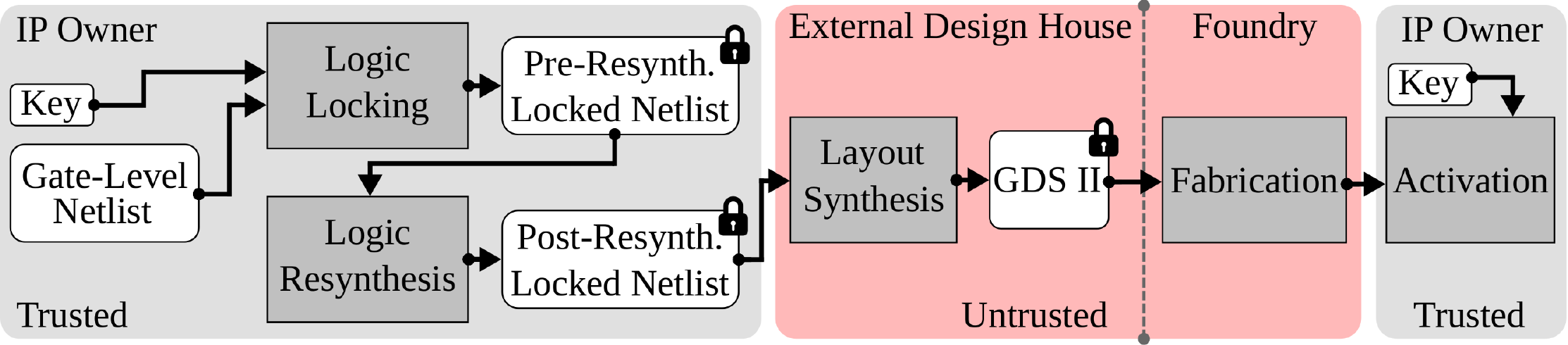}
	} 
	\caption{(a) Example Logic Locking and (b) Logic Locking in the IC Design and Fabrication Flow}
	\label{fig:LL}
\end{figure}

Logic locking introduces a secret key into the IC design and fabrication flow, as presented in Fig.~\ref{fig:LL}~(b). The IP owner (trusted regime) generates a gate-level netlist and uses a secret key to lock the design with a selected locking scheme. Afterwards, the locked design proceeds in the untrusted regime. This typically includes an external design house (for layout generation) and the foundry. After fabrication, the chip is returned to the owner for activation. Since the secret key is only known to the IP owner, logic locking protects the design while in the untrusted regime.

Note that the external design house is sometimes not included in the flow since some IP owners have internal front end (RTL and logic synthesis) and back end (layout generation) teams. However, especially smaller companies often subcontract parts of the design services, e.g., back end, to an external design house due to tight deadlines, cost reduction and the lack of necessary skills~\cite{sisejkovic2020scopes}. Moreover, the secret key induced by the locking mechanism is not required for a successful layout design and chip production.

\subsection{Terminology}\label{preliminaries:nomen}
The mentioned IC design flow includes the usage of netlists in different formats. In this work, we often use the term \textit{resynthesized} netlist. A netlist is regarded as resynthesized as it has to be synthesized \textit{once more} after logic locking is applied to maximize the structural changes induced by a locking scheme. 
Therefore, it is important to differentiate a \textit{pre-resynthesized} and a \textit{post-resynthesized} netlist. As shown in Fig.~\ref{fig:LL}~(b), a locked pre-resynthesized netlist is a netlist which has not been resynthesized after logic locking is applied. A locked post-resynthesized netlist is locked and resynthesized. In the rest of the paper, we refer to a gate-level netlist as \textit{netlist}.

In this work, schemes that are resilient against key-guessing (prediction) attacks of a machine learning model, are referred to as ML-resilient. In the general case, schemes that are resilient against guessing attacks performed by any system (human or artificial), are referred to as learning-resilient. In this context, ML resiliency is a subset of learning resiliency.

\subsection{Attack Model}
As previously discussed, the effectiveness and applicability of different logic locking schemes strongly depend on the assumed attack model. In this work, we assume the most restrictive attack model from the attacker's point of view: the adversary has \textit{only} access to the locked netlist (directly as a rogue external designer or by reverse-engineering the GDSII in the foundry). Since an activated IC is not available, a comparison to golden I/O patterns is not possible. By definition, this attack model forces an attacker to recover the key based on the least amount of information available.

Furthermore, it is assumed that the locations of the key inputs (pins) are known to the adversary. Only the key itself remains a secret. These are commonly accepted assumptions regardless of the availability of an oracle IC~\cite{comparativeAnalysis, evo2017, decadeOfLocking}.

\subsection{Deep Learning and Neural Networks} \label{CNNinfo}
Artificial neural networks represent a general class of machine learning algorithms. In this work, we exploit the capabilities of convolutional neural networks; one of the most significant networks in the landscape of deep learning techniques~\cite{cnnSurvey2020}. CNNs are a specific type of neural networks which were initially designed for 2-dimensional convolutions, inspired by the biological process of the visual cortex of animals~\cite{LeCun1998}. Since their introduction, CNNs have successfully pioneered the field of object classification and detection in the spectrum of deep learning. Their achievements are further fueled by the increasing availability of fast computation and vast amounts of data. The key enabling factor for their superior performance is their ability to extract meaningful features by exploiting hierarchical data patterns, i.e., recognizing spatial and temporal \textit{correlation in data}. Moreover, CNNs are well suited for object detection in unstructured data such as images. 

\begin{figure}[t]
	\centering
	\includegraphics[width=\linewidth]{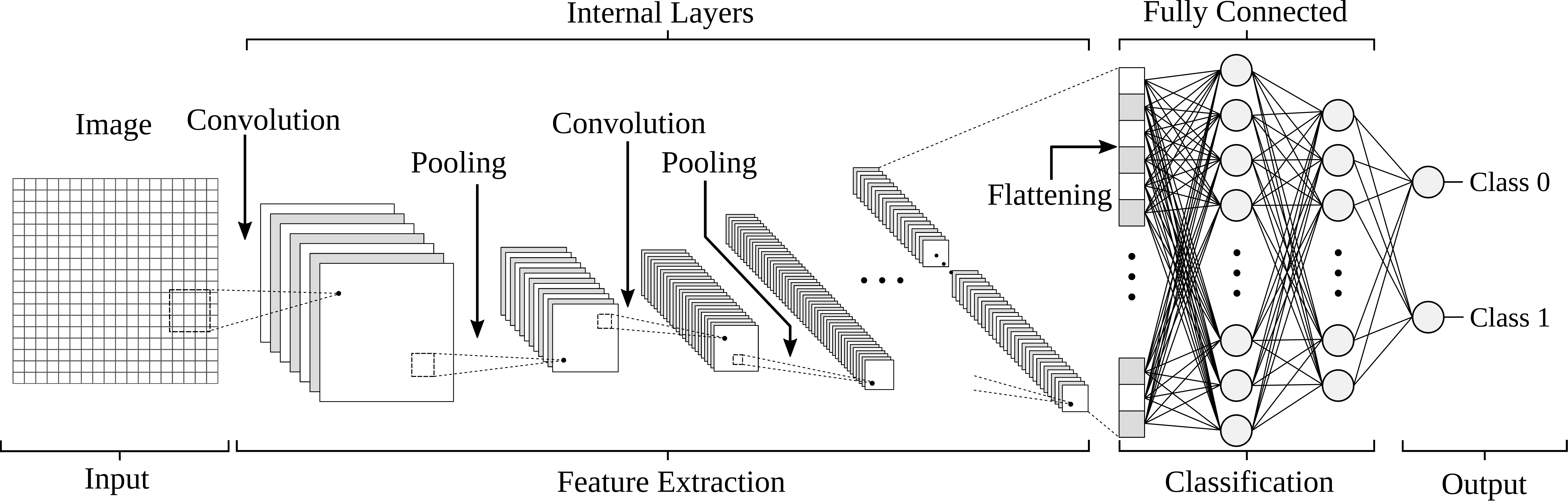}
	\caption{Architecture of a Convolutional Neural Network}
	\label{fig:cnn-blueprint}
\end{figure}

For this work, it is important to understand the very basic structure of a CNN. A general CNN architecture blueprint is presented in Fig.~\ref{fig:cnn-blueprint}. The input to the network is an image. The output is typically a classification prediction (e.g., whether the key bit value is 0 or 1). The two major components of the network include the \textit{feature extraction} and the \textit{classification}. The feature extraction is based on the typical CNN layers that we refer to as \textit{internal layers} (sometimes also referred to as hidden layers). The classification part is built using a fully connected neural network. The term "fully connected" implies that every neuron in the previous layer is connected to every neuron of the next layer. In general terms, the fully connected layers are trained to \textit{classify} images based on the \textit{features} extracted using the internal layers. The internal layers are typically a sequence of two layer types: convolution and pooling. 

Convolution layers extract features from the input image by detecting local conjunctions of features from previous layers. The result of the convolution is typically further processed using an activation function to increase the non-linearity. A commonly used activation function is the Rectified Linear Unit (ReLU). Furthermore, pooling layers perform the down sampling of feature maps, thereby progressively reducing the spatial size of the representation to reduce the amount of data and computation in the network. The most common approach used in pooling is known as max pooling.

After the feature extraction, the flattening layer converts the data into a 1-dimensional array that is fed as input to the fully connected layers. These layers assemble a traditional multi-layer network that use a softmax activation function in the final output layer. As mentioned before, the internal layers extract high-level features of the input image. Based on these features, the fully connected layers classify the input image into various classes. More details can be found in~\cite{guideToCNNs2018,LeCun1998}.

Another important type of ANN is known as Multi-Layer Perceptron (MLP). An MLP is a feedforward neural network which consists of multiple layers. An MLP takes flattened vectors as inputs rather than 2D or 3D data (similar to the classification part in Fig.~\ref{fig:cnn-blueprint}~(b)). In this work, we apply a simple MLP with fully-connected layers and compare the results to the more complex CNNs. Even though MLPs are nowadays succeeded by the more efficient CNNs, they are still a popular choice in machine learning for a variety of classification tasks.

\textbf{Motivation for ANNs:} In this work, it turns out that the problem of predicting a correct key bit can be represented in the form of images that are suitable for processing with ANNs. The images show a clear correlation between key values and recognizable netlist structures, especially offering the potential for feature extraction using CNNs. Further justification is presented in Section~\ref{CNNjustification} after introducing all relevant concepts.

\subsection{Genetic Algorithms and Neuroevolution}\label{ga-info}
Genetic Algorithms (GAs) belong to the class of evolutionary algorithms that represent population-based meta-heuristic optimization techniques inspired by biological evolution~\cite{Eiben2015}. The solutions in a population compete based on their quality (fitness) that is evaluated using a fitness function. Throughout multiple generations, solutions with a higher fitness are evolved by reapplying the genetic operators (selection, crossover and mutation) to the current population. A key component of GAs is the specific representation (encoding) of a solution as presented to the evolutionary algorithm. This \textit{manipulable} representation is referred to as \textit{genotype}. The actual solution to a problem, i.e., manifestation of the genotype is known as \textit{phenotype}. An important feature of evolutionary algorithms is their ability to work with black-box optimization problems, i.e., the algorithms do not need any domain-specific knowledge to perform the optimization.

The evolution of neural networks by using evolutionary algorithms is referred to as \textit{neuroevolution}~\cite{neuroevolution2008}. In this case, the task of the evolutionary process is to automatically evolve CNN architectures that are suitable for a selected problem domain. In general, neuroevolution can be used to evolve the network weights, structure or both. In this work, the genotype denotes a series of rules for constructing the CNN, while the phenotype represents a fully constructed CNN architecture. Note that neuroevolution is only one of many Neural Architectural Search (NAS) methods available~\cite{NASSurvey2018}. NAS belongs to AutoML; a research area that focuses on methods to make machine learning more efficient and automated. More details are available in~\cite{AutoMLSurvey2020}.

\textbf{Motivation for neuroevolution:} The reason for using neuroevolution in this work is twofold. First, selecting a suitable CNN architecture used for feature extraction is a challenging task, since it depends on the problem domain itself and often includes optimizing different hyperparameters. By using GAs to automatically evolve the architecture, the neuroevolutionary approach facilitates the exploration and selection of CNNs that are specifically drafted for the given problem domain. Hereby, the evolution is focused on resolving the issue of finding a suitable number and type of internal layers in the feature extraction component of the CNN. The reason why we focus on this aspect of the architecture is that all other components can be set to standard values which are often used in literature, while the selection of the internal layers highly depends on the problem domain. Secondly, the evolution process drastically speeds up the architecture selection as we assume a limited amount of time and resources to perform the task. Therefore, iterating through all possible CNN architectures is not a viable option. Interestingly, this showcases an important result: a potential attacker is able to efficiently attack locking schemes with SnapShot even without specialized knowledge about designing CNN architectures in a limited amount of time and with limited resources. Finally, by focusing on the architectural aspect of CNNs, we are able to learn about the complexity of the underlying classification problem based on the evolved networks (discussed in Section~\ref{res:networks}). Note that neuroevolution can be used for simple MLPs as well. However, since MLPs tend to have only a few layers, we design and evaluate the MLP architectures manually.

\section{Related Work} \label{related}
Only a few oracle-less attacks have been reported so far. The first is the desynthesis attack~\cite{desynthesisAttack2017}. Based on the observation that the locked and the original netlist should be similar in terms of the number and type of gates, the attack resynthesizes the netlist with a random key. Afterwards, the attack uses Hill Climbing search to find a key that yields a maximum similarity between the locked and resynthesized netlist. However, this attack is not scalable for larger key lengths and presumes a sound knowledge of the synthesis tool. Therefore, its efficiency for large-scale designs remains uncertain. In comparison, SnapShot scales linearly with the key length.

The recently introduced redundancy attack reveals a novel vulnerability of logic locking schemes based on identifying logic redundancies introduced by incorrect key bits~\cite{redundancyIdentification2019}. The attack prunes out the incorrect values of a key when it introduces a significant level of functional redundancy. 
However, a recent work has devised a locking scheme that is resilient against this attack~\cite{redundancyShield2019}. Therefore, we do not conduct a comparison to this attack. Note that SnapShot is also applicable to the mentioned resilient locking since it is based on key gate insertion.

Another oracle-less attack is known as SWEEP~\cite{sweepAttack2019}. This attack exploits the structural changes induced during the synthesis process to extract correct key values. It works as follows. The attack hard-codes both the correct and the incorrect key bit value for each key input based on a training set of locked benchmarks. For each value of each key input, the attack reoptimizes the design and performs a synthesis analysis. Based on the synthesis report, the attack identifies any design features that are correlated to the correct key values. Finally, the same procedure is repeated for the target netlist. Based on a comparison using a scoring algorithm to evaluate each design feature, the attack identifies the correct key bits with a high confidence. However, this attack is not applicable to XOR/XNOR-based locking, therefore we do not perform a comparison to this attack.

The recent oracle-less Topology-Guided Attack (TGA) is based on the intuition that basic functions in a logic cone are typically repeated multiple times in a netlist~\cite{TGA2019}. These are referred to as Unit Functions (UFs). In case a key gate is placed in an instance of a UF, the correct key bit value can be recovered by searching through equivalent UFs that are constructed using all hypothesis key values by self-referencing the given netlist. This attack is similar to our work in the sense that it exploits the fact that limited XOR/XNOR-induced changes are mostly same for similar portions of a netlist. To prevent TGA, the authors present a scheme in which either a key gate is placed in a unique UF or all equivalent UFs are locked simultaneously. Both result in the TGA search being unsuccessful. Note that even if this scheme is used, it does not affect the way a single key gate is inserted. Therefore, the solution still exhibits the same vulnerabilities in terms of SnapShot.

\begin{figure}[t]
	\centering
	\includegraphics[scale=0.53]{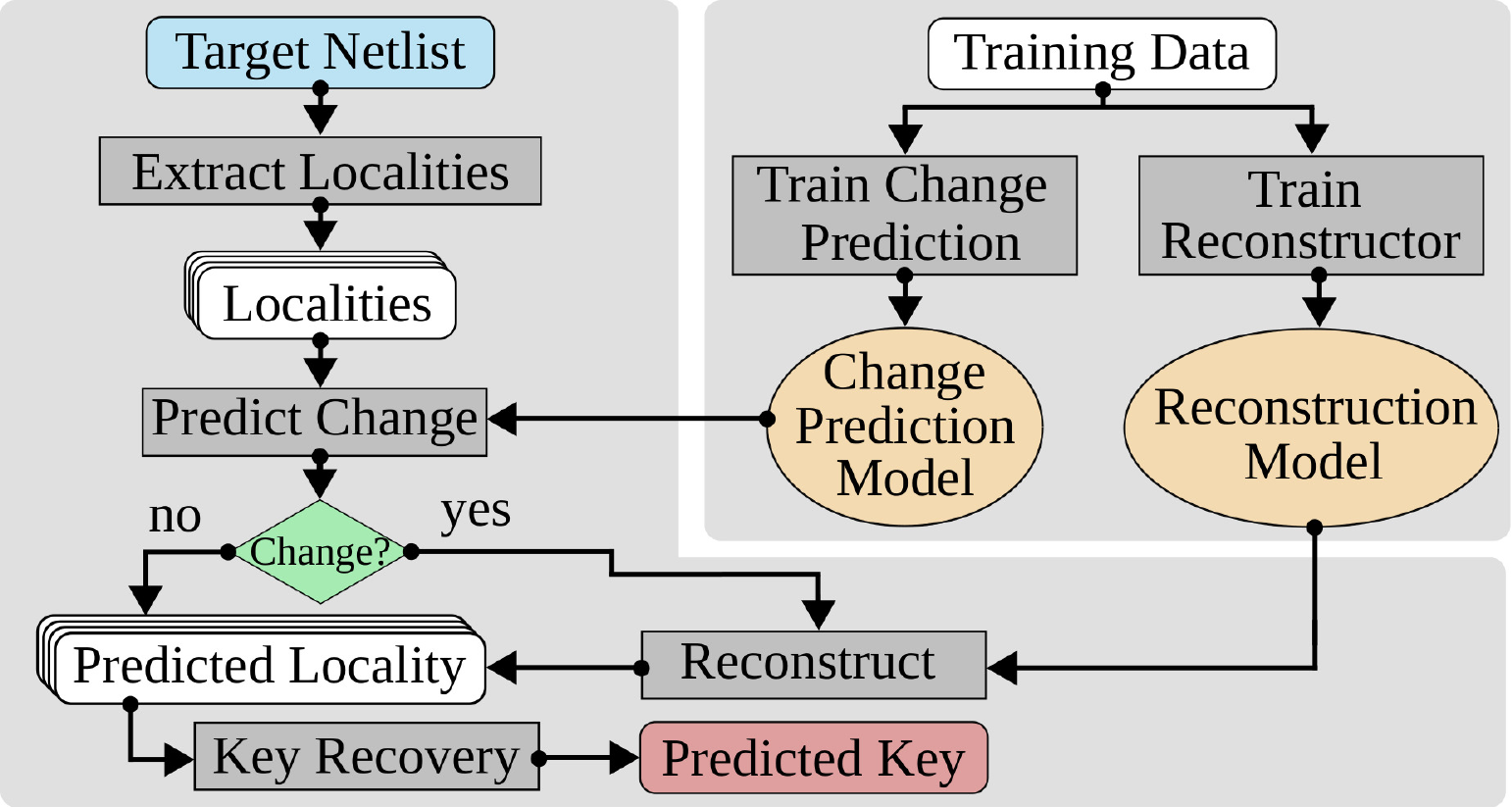}
	\caption{SAIL Attack Flow}
	\label{fig:sail}
\end{figure}

In recent times, several ML and Evolutionary Computation (EC) based attacks have emerged. The BOCANet attack utilizes deep recurrent neural networks to compromise hardware obfuscation~\cite{BOCANet2019}. GenUnlock formulates logic locking as an optimization problem and utilizes genetic algorithms to perform an attack~\cite{GenUnlock2019}. Similarly, the particle swarm optimization guided attack uses EC to find an approximate key that produces correct outputs in most cases~\cite{particleSwarmBasedLL2020}. Even though these attacks utilize similar algorithmic models as SnapShot, they all depend on having access to input/output observations taken from an activated IC. Therefore, these do not fall into the oracle-less category.

The most relevant attack for a comparison with SnapShot is SAIL~\cite{SAIL2019}. SAIL is a structural attack on logic locking that utilizes ML algorithms to retrieve local logic structures from the locked post-resynthesized netlist (referred to as \textit{target netlist}). The applied algorithms include a support vector machine, random forest and a multi-layer neural network. 
The SAIL attack flow is presented in Fig.~\ref{fig:sail}.
The attack strategy is as follows. Pre- and post-resynthesized locked designs are provided as training data to train two different models: the Change Prediction Model (CPM) and the Reconstruction Model (RM). Given a netlist subgraph (locality) extracted near the selected key input, i.e., key gate, CPM predicts whether a structural change has occurred, i.e, whether the synthesis tool has induces any changes around the inserted key gates. If a change is predicted, RM is utilized to predict the pre-resynthesized subgraph, i.e., it reverses the changes of the logic synthesis. Out of the reconstructed subgraph, the key value is determined \textit{based on the properties} of XOR-based locking: an XOR gate is inserted for key bit 0, and XNOR for a key bit 1~\cite{SAIL2019, SURF2019}.

\textbf{Advantages of SnapShot:} To understand the advantages of SnapShot, we need to analyze the limitations of SAIL. The major comparison points are as follows:
\begin{itemize}
	\item SAIL is \textit{only} applicable to XOR-based locking, since it relies on its inherent security properties (XOR bound to 0 and XNOR bound to 1). Therefore, it is \textit{not possible} to utilize SAIL to predict a key for schemes that do not rely on the mentioned XOR/XNOR property, such as a MUX-based locking scheme. In comparison, due to the nature of the attack and the customizable representation, SnapShot can be applied to a wide range of schemes and netlist structures. Note that the actual prediction accuracy depends on the security foundation of the evaluated locking schemes, yet the applicability is ensured. 
	\item In essence, SAIL learns to reconstruct the synthesis changes to predict the original netlist structure. Therefore, the key prediction in SAIL is only possible for reconstructed or unchanged netlist localities. This implies that the success of SAIL heavily depends on the complexity of the synthesis procedure that is used after logic locking is applied. Consequently, the attack success depends on having access to the original synthesis tool that was used in the design flow of the netlist under attack, thereby introducing another obstacle for the adversary. In comparison, SnapShot allows a \textit{direct} prediction of the key bit values based on the provided netlist localities, since the attack is using only post-resynthesized netlists. Therefore, the direct prediction enables a simpler and more flexible attack vector for an adversary.
	\item Finally, SAIL depends on training two different models (CPM and RM), while SnapShot includes the training of only one model, thereby offering another advantage for an adversary. Even with a simplified learning model, SnapShot achieves a significant improvement in the key prediction accuracy.
\end{itemize}

\section{The SnapShot Attack} \label{snapshot}
The objective of SnapShot is the immediate prediction of a key bit value based on the extracted \textit{locality} (subgraph of gates) around each key gate. The reasoning to this attack is as follows. As per attack model, the adversary is aware of the exact locations of all key inputs. For each key input, the exact key gate (or structure) can be identified by simply following the key input signal. This can be repeated for each bit of the key individually. For XOR/XNOR locking schemes, the key gate is the first gate that is connected to a given key input. Since current gate insertion-based locking schemes only induce \textit{limited local changes} to the netlist~\cite{SAIL2019}, it is possible for the underlying neural network to predict which key bit value is expected for a selected locality around a specific key gate. 

The SnapShot attack flow is presented in Fig.~\ref{fig:snapshot}. The input of the flow includes a target netlist, the number of samples used for the data generation ($N$) and the key length ($K$). The target netlist is a locked and resynthesized netlist (see Section~\ref{preliminaries:nomen} for details). The final output of the attack is the predicted activation key for the target netlist. The flow consists of four stages: setup, extraction, ANN design and deployment. In the following, all steps are discussed in detail in the given order, thereby referencing the flow in Fig.~\ref{fig:snapshot}.

\subsection{Setup} \label{snapshot:setup}
\begin{figure}[!t]
	\centering
	\includegraphics[width=\columnwidth]{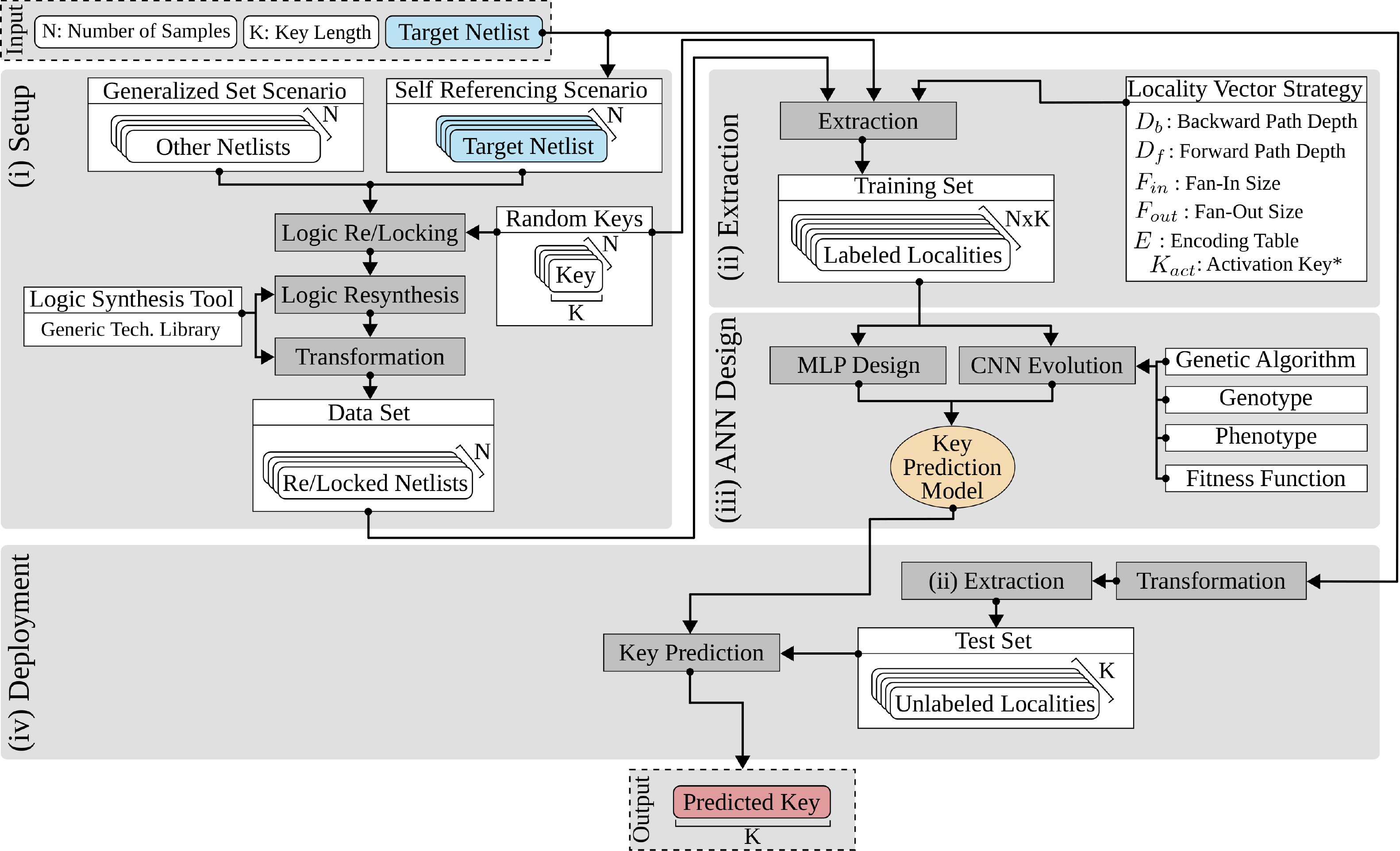}
	\caption{SnapShot Attack Flow}
	\label{fig:snapshot}
\end{figure}

The setup includes the preparation of a \textit{data set} upon which a \textit{training set} is extracted in the next stage (Fig.~\ref{fig:snapshot}~(step $i$)). First, it is important to understand the difference between these two sets. The data set represents a set of locked benchmark netlists, i.e., a set of post-resynthesized netlists locked with a selected locking scheme. The training set denotes the actual labeled localities in the form of vectors that are extracted from the data set for each key input of every benchmark (explained in the next stage). The setup can be configured to target \textit{two realistic attack scenarios}: the Generalized Set Scenario (GSS) and the Self-Referencing Scenario (SRS).

\textbf{GSS:} In GSS, the adversary tries to predict the correct key for a locked target netlist using an ML model that is trained based on the extracted data from a \textit{generalized data set}. The generalized data set is generated by locking a large number of \textit{other} (different from target) netlists with random keys with the selected scheme. Note that the generalized data set does not contain the target netlist and that the netlists can implement any functionality. Each netlist, out of the $N$ selected netlists in the generalized set, is locked with a single key of length $K$. After locking, all locked netlists are resynthesized, resulting in $N$ locked and resynthesized netlists. These are used to train the model and attack the target. \textit{The reasoning behind GSS is that in a real-life attack, an adversary can use any available circuits (e.g., open-source or in-house designs) to train the model.} To the best of our knowledge, this scenario has not been addressed in existing literature so far.

\textbf{SRS:} In SRS, instead of using other netlists, the adversary can create a data set by \textit{relocking}, i.e., self-referencing, the locked target netlist with additional keys. This is performed by copying the target netlist $N$ times and relocking all $N$ copies with separate keys of length $K$. Finally, the $N$ locked netlists are resynthesized.
SRS implies that each relocked netlist has two sets of key inputs: the \textit{original} target key inputs under attack that are same for every netlist and the \textit{additional} key inputs from the relocking phase that are different for every netlist. Afterwards, the additional key inputs of the generated set are used to train the ML model to predict the key of the target netlist. In both attack scenarios, the training is done based on a set of locked netlists. However, in GSS the set consists of a variety of different designs, while in SRS it consists of many copies of the target netlist. Nevertheless, SRS has the advantage that an adversary does not need to possess other netlists to train the ML model and that the training is based on data that is closely related to the actual target. The latter has a positive impact on the prediction capabilities of SnapShot, as discussed in Section~\ref{results}. 

\begin{figure}[t]
	\centering
	\includegraphics[width=0.7\columnwidth]{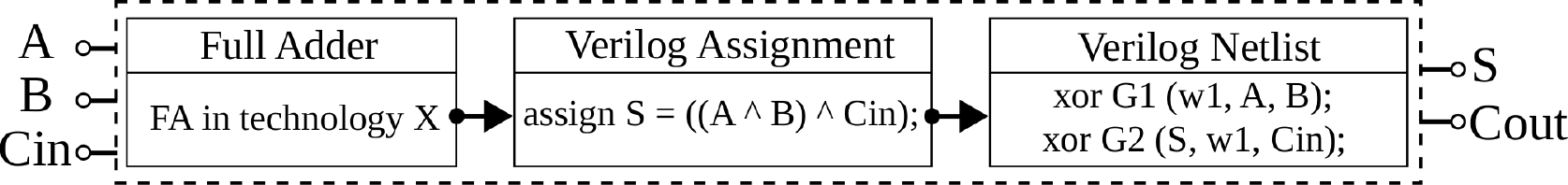}
	\caption{Example: Netlist Transformation}
	\label{fig:tac}
\end{figure}

\textbf{Transformation:}~In both GSS and SRS, we perform one essential step after the resynthesis: the netlist transformation. This step ensure the translation of the synthesized netlists into a generic technology-independent format. We perform this step by instructing the underlying synthesis tool to map the design to a generic technology library and store it using plain Verilog. When storing, the synthesis tool transforms each technology cell into its equivalent Boolean function. This function is represented using simple combinational \textit{assignments} in Verilog. These assignments are parsed, transformed into three-address code and finally stored using primitive Verilog gates. An example transformation for the sum operation (output S) of a Full Adder (FA) is shown in Fig.~\ref{fig:tac}. First, the output S of the technology-dependent FA is written out as a Verilog assignment. Finally, the Verilog expression is transformed into primitive Verilog gates. Therefore, the resulting file is a technology-independent gate-level netlist. 

For logic synthesis and mapping, we utilize the Synopsys Design Compiler and its accompanying generic technology library. Being technology-independent has the advantage of not specializing the attack for a particular technology node, since using a specific library can negatively influence the generalization of the approach. Moreover, an adversary can always resynthesize a design with a generic library regardless of the underlying technology.

The transformation step offers one more feature. Since the Verilog assignments can be processed in the form of three-address code, the transformation step enforces that every gate in the resulting netlist has \textit{a maximum of two inputs}. 
Only BUF and INV gates have one input. This property is further utilized in the extraction stage of SnapShot.

\begin{table}[!b]\footnotesize
	\caption{Locality Vector Encoding}
	\centering
	\tabcolsep=0.13cm
	\begin{tabular}{cccccccccc}\hline
		\textbf{Gate Type} & NOT & AND & NAND & OR & XOR & NOR & XNOR & BUF & FF \\\hline
	\rowcolor{gray!20}	\textbf{Code} &  1 & 2 &3&4&5&6&7&8&9  \\\hline
	\end{tabular}
	\label{tab:encoding}
\end{table}

\subsection{Extraction} \label{snapshot:extraction}
Once the data set is prepared, the next step is to assemble the training set (Fig.~\ref{fig:snapshot}~(step $ii$)). This set consists of a large number of \textit{labeled localities}. Remember that a locality is simply a subgraph around a selected key gate that is connected to a particular (1-bit) key input. The extraction of the localities is not uniquely defined, i.e., a variety of \textit{extraction strategies} are viable. However, a selected strategy must ensure that all necessary information that is crucial for the ANN learning and prediction process is captured in the extracted representation of a locality. This information includes the number of gates, the gate types and their relationship (how the gates are connected).

\textbf{Locality Vector:}~As specifically CNNs work well with unstructured image data, it is advantageous to represent all subgraphs in a form that can be translated into images. Hereby, the prediction problem is represented as the \textit{classification} of the extracted image (locality) as key bit 0 or 1. For this purpose, we propose a \textit{locality vector} representation. A locality vector is a sequence of \textit{numbers} representing the extracted subgraph. The vector reflects the actual netlist subgraph in the following way. Each entry in the vector represents a single location in the netlist. A location can be filled by a single gate or remain empty (no gate present). The actual value of the entry simultaneously determines two properties of a location: the existence of a gate on that location and the gate type. These properties are defined by a selected encoding of the gate types. We selected the encoding in Table~\ref{tab:encoding}. Note that this encoding includes all possible primitive gate types that can occur in a design. A particular value uniquely identifies the gate type. The existence of a gate is encoded indirectly: an empty location is marked with the value 0. If the value is greater than 0, the location is filled by a gate of a particular type.
\begin{algorithm}[t]\footnotesize
	\caption{LVE: \textbf{L}ocality \textbf{V}ector \textbf{E}xtraction}
	\label{alg:LVE}
	\begin{algorithmic}[1]
		\Require{Locked Generic Netlist ($Net$), Backward Path Depth ($D_{b}$), Forward Path Depth ($D_{f}$), Fan-In ($F_{in}$), Fan-Out ($F_{out}$), Encoding Table ($E$) and Activation Key (optional) ($K_{act}$)}
		\Ensure{Set of Locality Vectors ($L$)}
		\State $L \gets [\emptyset]$\Comment{Prepare locality vector set}
		\State $K \gets \abs{Net.K_{in}}$\Comment{Length of key input}
		\State $T \gets \{Backward, Forward\}$\Comment{BFS type}
		\item[]
		\For{$i = 0$ to $K$}
		\State $KG_{i} \gets$ GateConnectedTo($Net.K_{in}$[i])
		\State $G_{i} \gets$ InputGatesOf($KG_{i}$)[1]
		\item[]
		\LineComment{Extract values}
		\State$l_{b} \gets$ BFS($T.Backward, G_{i}, E, D_{b}, F_{in}$)
		\State$l_{kg} \gets E$[Type($KG_{i}$)]
		\State$l_{f} \gets$ BFS($T.Forward, KG_{i}, E, D_{f}, F_{out}$)
		
		\item[]
		\LineComment{Merge values}
		\State $L[i] \gets \{l_{b}, l_{kg}, l_{f}\}$
		
		\item[]
		\LineComment{If $K_{act}$ given: label the locality}
		\If{$K_{act}~!=~\{\emptyset\}$}
		\State $L[i] \gets \{K_{act}[i], L[i]\}$
		\EndIf
		\EndFor
		\State \textbf{return} $L$
	\end{algorithmic}
\end{algorithm}

\textbf{Extraction Procedure}:~So far, we have discussed how the locality vector is able to store information about the existence of particular gates as well as their types. The next step is to understand how the relationship of the gates is stored. The way the gates are connected is reflected by the \textit{position} of each value in the vector. This means that based on the location of a value in the vector, we can determine its input and output gates. As before, encoding the relationship in the position is not uniquely defined. For this purpose, we define the Locality Vector Extraction (LVE) procedure shown in Algorithm~\ref{alg:LVE}. LVE works as follows. The main LVE loop repeats for every (1-bit) key input (line 4). In the first step (line 5), LVE finds the first gate ($KG_{i}$) connected to the currently observed key input ($Net.K_{in}[i]$), as well as the first non-key input connected to $KG_{i}$ ($G_{i}$, line 6). In the second step, LVE extracts three independent sections of a locality vector. The first section is extracted by performing a Breadth-First Search (BFS) starting from $G_{i}$ towards the netlist inputs (backward path, line 8). The backward BFS starts from $G_{i}$ since we already know that the first input of $KG_{i}$ is a key input. The second section extracts the encoding of the "center" gate $KG_{i}$ by looking up the encoding table $E$ (line 9). Note that $E$ is defined in Table~\ref{tab:encoding}. The third section is extracted by performing a BFS starting from $KG_{i}$ in the direction of the netlist outputs (forward path, line 10). Finally, all three sections are concatenated to form a locality vector (line 12). In case an activation key ($K_{act}$) is provided, each locality is labeled with its respective activation key-bit by prepending the value to the locality vector (lines 14-16). $K_{act}$ is an optional argument since LVE can be used to generate a training set (labeled vectors) as well as a test set (unlabeled vectors).

The BFS used in LVE for the extraction of the backward and the forward path is shown in Algorithm~\ref{alg:BFS}. The BFS receives the following inputs: the BFS type ($T_{bfs}$) defining whether a backward or forward search is performed, the root gate ($R$), the encoding table ($E$), the maximum allowed search depth for the backward ($D_{b}$) and forward ($D_{f}$) path, the fan-in ($F_{in}$) and fan-out ($F_{out}$) size per gate. Note that the depths and fan-sizes are each represented by one variable ($D$ and $F$). The algorithm itself is a standard BFS with variable properties as discussed in the following.

The search direction is selected by $T_{bfs}$. For a backward search, the neighbor gates are all \textit{input} gates of the currently processed gate $G_{i}$ (line 10). For a forward search, the neighbor gates are all \textit{output} gates of $G_{i}$ (line 12). The algorithm assumes a fixed number of possible inputs (for backward BFS) or outputs (for forward BFS) per gate. This property is defined by $F$. As the transformation in the setup stage of SnapShot ensures that each gate has a maximum of two inputs (see Section~\ref{snapshot:setup}), for a backward BFS, $F_{in}$ is set to 2 (LVE line 8). However, a single gate can have any number of possible output gates connected to it. Therefore, for a forward BFS, the number of output gates is defined by $F_{out}$ (LVE line 10). Fixing the number of inputs/outputs ensures that each locality has the same length regardless of the actual netlist subgraph it represents. In case a single gate deviates from the fixed value, the BFS algorithm creates additional \textit{empty} gates (lines 15-17). The encoding for an empty gate is the value 0. Finally, the BFS terminates when the selected search depth is reached (lines 34-36).

The actual locality vector $l$ is assembled by extracting the encoding for each visited neighbor (lines 25-29). Hereby, the encoding of the currently processed neighbor $N_{j}$ is either prepended (for backward BFS) or appended (for forward BFS) to $l$. The reason for using different insertion methods depending on $T_{bfs}$ is to reflect the topological order of the gates in the locality vector. If we assume that the key gate is placed in a central position in $l$, then gates that are closer to the netlist inputs are placed towards the left end of $l$. Consequently, gates that are closer to the netlist outputs are placed towards the right end of $l$.

\begin{figure}[!t]
	\centering
	\includegraphics[width=\columnwidth]{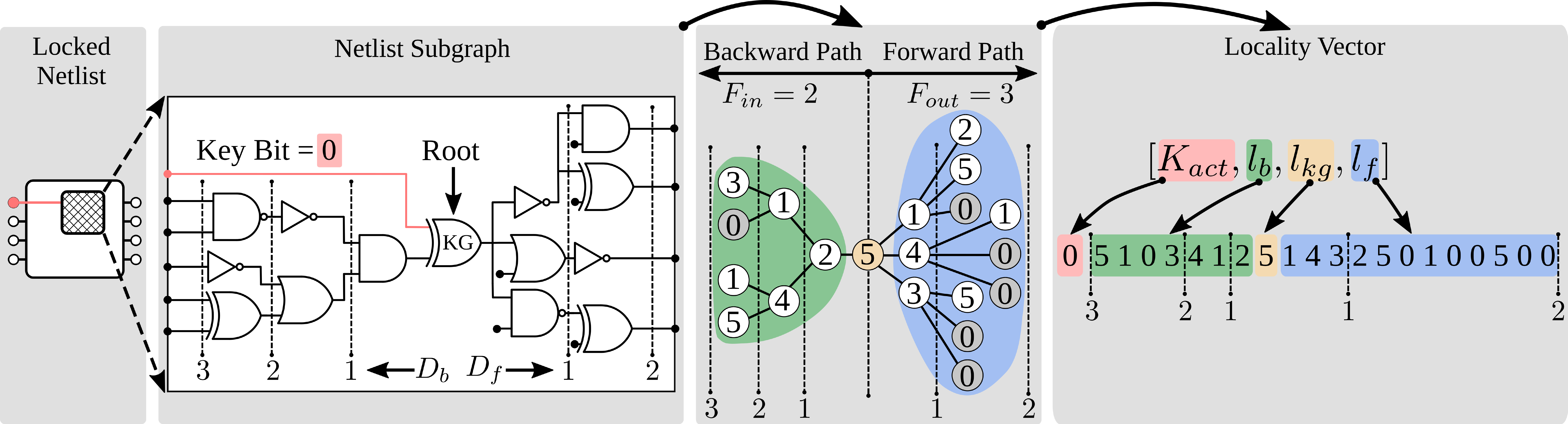}
	\caption{Example: Localities Extraction and Representation}
	\label{fig:representation}
\end{figure}

\begin{algorithm}[!htbp]\footnotesize
	\caption{BFS Extraction Procedure}
	\label{alg:BFS}
	\begin{algorithmic}[1]
		\Require{BFS Type ($T_{bfs}$), Root Node ($R$), Encoding Table ($E$), Max Depth ($D$) and Fan-In/Out ($F$)}
		\Ensure{Locality Vector ($l$)}
		\State $l \gets \{\emptyset\};~Q \gets [\emptyset]$\Comment{Prepare locality vector and FIFO queue}
		\State $done \gets FALSE; D_{i} \gets 1$\Comment{Prepare control variable and set current depth}
		\State $Q.$Enqueue(R)\Comment{Add root gate}
		\State $R.$Visited$ \gets TRUE$\Comment{Set root as visited}
		\item[]
		\While{$!$IsEmpty($Q$)~\&\&~$!done$}
		\State $G_{i} \gets Q.$Dequeue()
		\State $N \gets [\emptyset]$
		\item[]
		\LineComment{Returns \{$\emptyset$\} for empty gate \& primary IOs}
		\If{$T_{bfs}$ == $T.$Backward}
		\State $N \gets$ InputGatesOf($G_{i}$)
		\Else
		\State $N \gets$ OutputGatesOf($G_{i}$)
		\EndIf
		
		\item[]
		\LineComment{Ensures: gate has $F$ ins or outs; Note: if $\abs{N} == F$, no empty gate is added}
		\For{$j=\abs{N}$ to $F$}
		\State $N[j] \gets$ NewEmptyGate()
		\EndFor
		
		\item[]
		\LineComment{Visit neighbors}
		\For{$j = 0$ to $F$}
		\State $N_{j} \gets N[j]$
		\If{$!N_{j}.$Visited}
		\State $Q.$Enqueue($N_{j}$)
		\State $N_{j}.$Visited$ \gets TRUE$
		\item[]
		\LineComment{Extract value}	
		\If{$T_{bfs}$ == $T.$Backward}
		\State $l.$Prepend($E$[Type$(N_{j}$)])
		\Else
		\State $l.$Append($E$[Type$(N_{j}$)])
		\EndIf				
		\EndIf
		\EndFor
		\item[]
		\LineComment{Check if max depth reached}
		\State $D_{i} \gets D_{i} + 1$
		\If{$D_{i}~==~ D$}
		\State $done \gets TRUE$
		\EndIf
		\EndWhile
		\State \textbf{return} $l$
	\end{algorithmic}
\end{algorithm}
One example extraction for a single 1-bit key input (marked red) is shown in Fig.~\ref{fig:representation}. The parameters are set as follows: $D_{b}=3$, $D_{f}=2$, $F_{in}=2$ and $F_{out}=3$. Using the encoding in Table~\ref{tab:encoding} and the proposed LVE procedure, one labeled locality vector is extracted in the format $[K_{act}, l_{b}, l_{kg}, l_{f}]$. The example also visualizes empty gates (marked with 0).

This representation has the following beneficial properties. It is complete in the sense that all types of subgraphs can be represented by traversing the netlist. The amount of data stored in one vector can be adapted by adjusting the \textit{depth} of the backward and forward path as well as the \textit{fan-in/out} size. Finally, the representation is expandable to sequential circuits by simply adding an encoding for a flip-flop. 

To verify that no information about the subgraph is lost, we can perform a simple test. By "reversing" the LVE procedure on a vector using the selected parameters, it is a trivial task to reconstruct the representing netlist subgraph. Therefore, it stands to reason that no information loss occurs. However, it is important to consistently follow the same deterministic steps in the extraction. Let us consider the $l_{b}$ section of the locality at $D_{b}=3$ in Fig.~\ref{fig:representation}. It is not relevant whether the extraction order is $l_{b}=[5, 1, 0, 3]$ or $l_{b}=[3, 0, 1, 5]$ as long as the same order is applied consistently for all sections and all localities. 

The proposed locality representation gives SnapShot one more advantage: universal applicability. For example, in MUX-based locking, a MUX is simply a collection of gates that can be represented with localities or the complete MUX can be encoded with a separate value. Moreover, the described setup and extraction stage of SnapShot can be used for any attack that exploits information from the netlist subgraphs. 

\subsection{ANN Design}\label{snapshot:evolution}
\begin{figure}[t]
	\centering
	\includegraphics[width=\columnwidth]{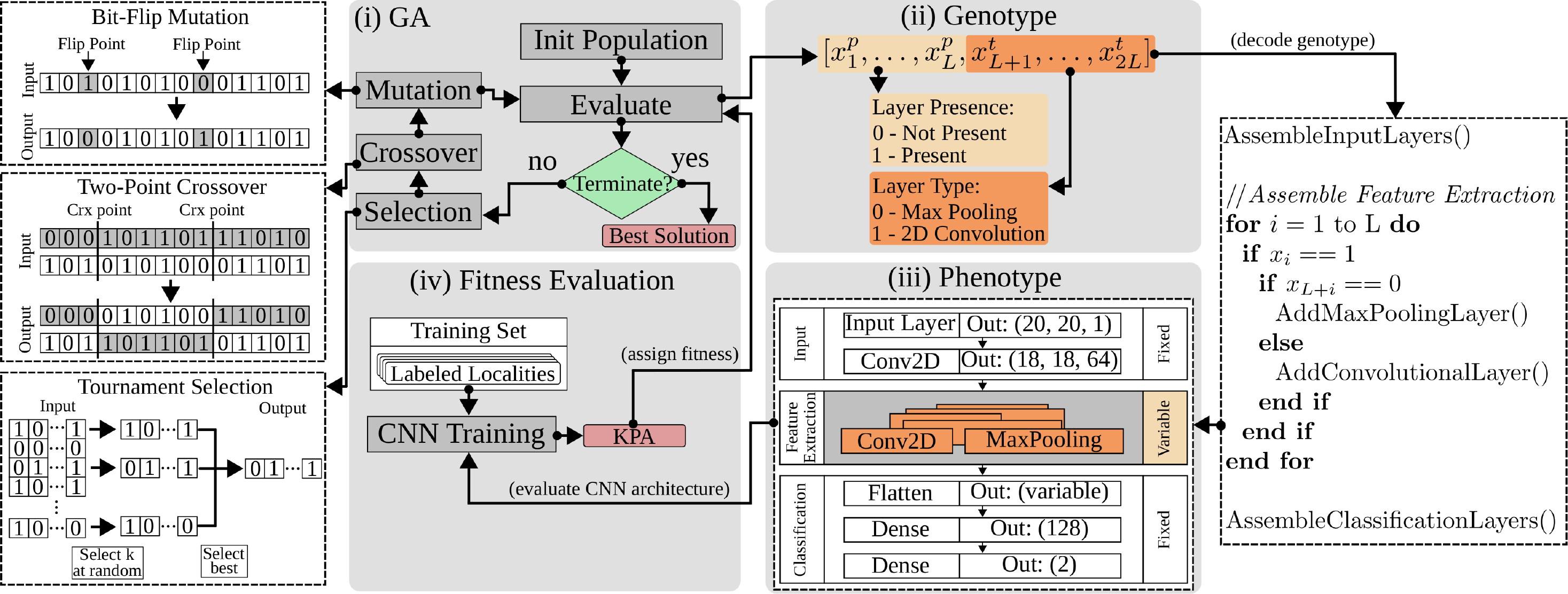}
	\caption{CNN Architecture Evolution with GA}
	\label{fig:ga}
\end{figure}

In this stage, a suitable ANN model is generated based on the prepared training set (Fig.~\ref{fig:snapshot}~(step $iii$)). In this work, we focus on two neural network types: simple fully connected MLPs and more complex CNNs. Note that any ML model can be used at this stage. Due to the relatively simple structure of MLPs, we define the suitable MLPs manually (Section~\ref{MLPModelSetup}). On the other hand, the architecture of CNNs is \textit{evolved} and \textit{trained} as discussed in the following. The CNN evolution contains four important components: the genetic algorithm, the genotype, the phenotype and the fitness function (see Fig.~\ref{fig:ga}).

\textbf{Genetic Algorithm:}~GAs are a suitable tool for black-box optimization problems. In SnapShot, the GA is utilized to automatically find a \textit{fitting CNN feature extraction architecture} for the given classification problem, thereby releasing the attacker from having to manually draft a suitable CNN. The GA internally works with a population of solutions to a specific problem. The solutions are adapted through the application of the genetic operators. Fig.~\ref{fig:ga} visualizes the exact operators (bit-flip mutation, two-point crossover and tournament selection) that are used in the implementation as well. This manipulation results in advancing the quality of the solutions through multiple generations (Fig.~\ref{fig:ga}~(step $i$)). To utilize a GA, it is necessary to define how a solution is represented (genotype), decoded (phenotype) and how it is evaluated (fitness function).

\textbf{Genotype:}~As explained in Section~\ref{CNNinfo}, a complete CNN architecture consists of two components: the feature extraction and the classification. These components are defined by a set of properties (e.g., number of layers, layer types, input/output dimensions and others). Some of these properties must remain \textit{fixed} to provide a foundation for valid CNN architectures. Therefore, we define a fixed frame that includes the following: the \textit{first} convolutional (input) layer and the \textit{last two} layers (a fully connected layer and the final output layer). Note that flattening is applied before the fully connected layer. This frame is based on common CNN architectural features, where the first internal layer is always a convolutional layer and the classification is typically assembled of two fully connected layers. On the other hand, \textit{the variable properties are the ones being evolved by the GA.} These include the \textit{number} and \textit{type} of internal layers. The type includes \textit{max pooling} and \textit{2D convolution}. These properties must be represented with a suitable genotype, i.e., the encoded representation of a concrete solution as presented to the GA. For SnapShot, the genotype is encoded as a \textit{bitstring} of size $2\cdot{L}$, where the first $L$ bits denote the presence of the individual internal layer ($[x^{p}_{1},\dots,x^{p}_{L}]$) and the last $L$ bits the layer types ($[x^{t}_{L+1},\dots,x^{t}_{2L}]$), as shown in Fig.~\ref{fig:ga}~(step $ii$). For example, if $x^{p}_{i}=1$, the $i$-th internal layer is present and its type is defined by $x^{t}_{i+L}$. Using this genotype, we fix the classification component and evolve the architecture of the feature extraction component.

\textbf{Phenotype:}~The discussed genotype is just a blueprint, i.e., a set of instructions defining how an architecture is assembled. Before a solution is evaluated, its genotype must be decoded into a concrete CNN instance, i.e., the phenotype. This is done by processing the genotype from left to right, thereby placing the existing internal layers \textit{between} the fixed layers according to the defined layer type, as presented in Fig.~\ref{fig:ga}~(step $iii$). The given input and output dimensions of the fixed layers are selected based on the input image size and the used hyperparameters (explained in Section~\ref{results}).

\textbf{Fitness Function:}~Once the genotype is decoded into a concrete CNN architecture, the fitness evaluation can be performed (Fig.~\ref{fig:ga}~(step $iv$)). The CNN is trained based on the training set prepared in the extraction stage of the SnapShot attack flow. A single CNN instance is trained to classify the training localities into 0 or 1 bit values for a selected amount of epochs. The number of epochs equals the number of times the complete training set is fed into the CNN. To evaluate the fitness, we define the \textit{performance measure} of the CNN prediction as the Key Prediction Accuracy (KPA). The KPA equals the percentage of correctly predicted key bits for a specific netlist, where each key bit input is predicted individually. For example, a KPA of 70\% indicates that 70\% of the bits of a single key are correctly predicted. This fitness function steers the evolution mechanism to evolve internal CNN layers which maximize the feature extraction process and result in higher classification accuracy for this particular problem.

Once the KPA for a particular CNN architecture is calculated for the training set, it is assigned to the respective solution in the GA population. After a selected number of generations, the GA terminates and returns the \textit{best solution} (architecture), i.e., the one with the highest KPA. This final solution is further used in the next stage of SnapShot.

\subsection{Deployment}
Once the ANN architectures are fixed, the ANN prediction model (the best available trained CNN architecture or defined MLP) is deployed to predict the key for the locked target netlist (Fig.~\ref{fig:snapshot}~(step $iv$)). To perform the prediction, the target netlist is transformed into a generic netlist (same procedure as described in Section~\ref{snapshot:setup}) and its localities are extracted into a \textit{test set}. The only difference to a training set is that now the localities are not labeled (since the correct key is unknown). Afterwards, each locality is presented to the prediction model to predict its key bit value, resulting in the final SnapShot output: \textit{the predicted key}. Since each key bit value can be predicted (attacked) independently, SnapShot is easily scalable to larger designs. 
\begin{figure}[!t]
	\centering
	\subfloat[0-Labeled Localities Image]{
		\includegraphics[width=0.35\columnwidth]{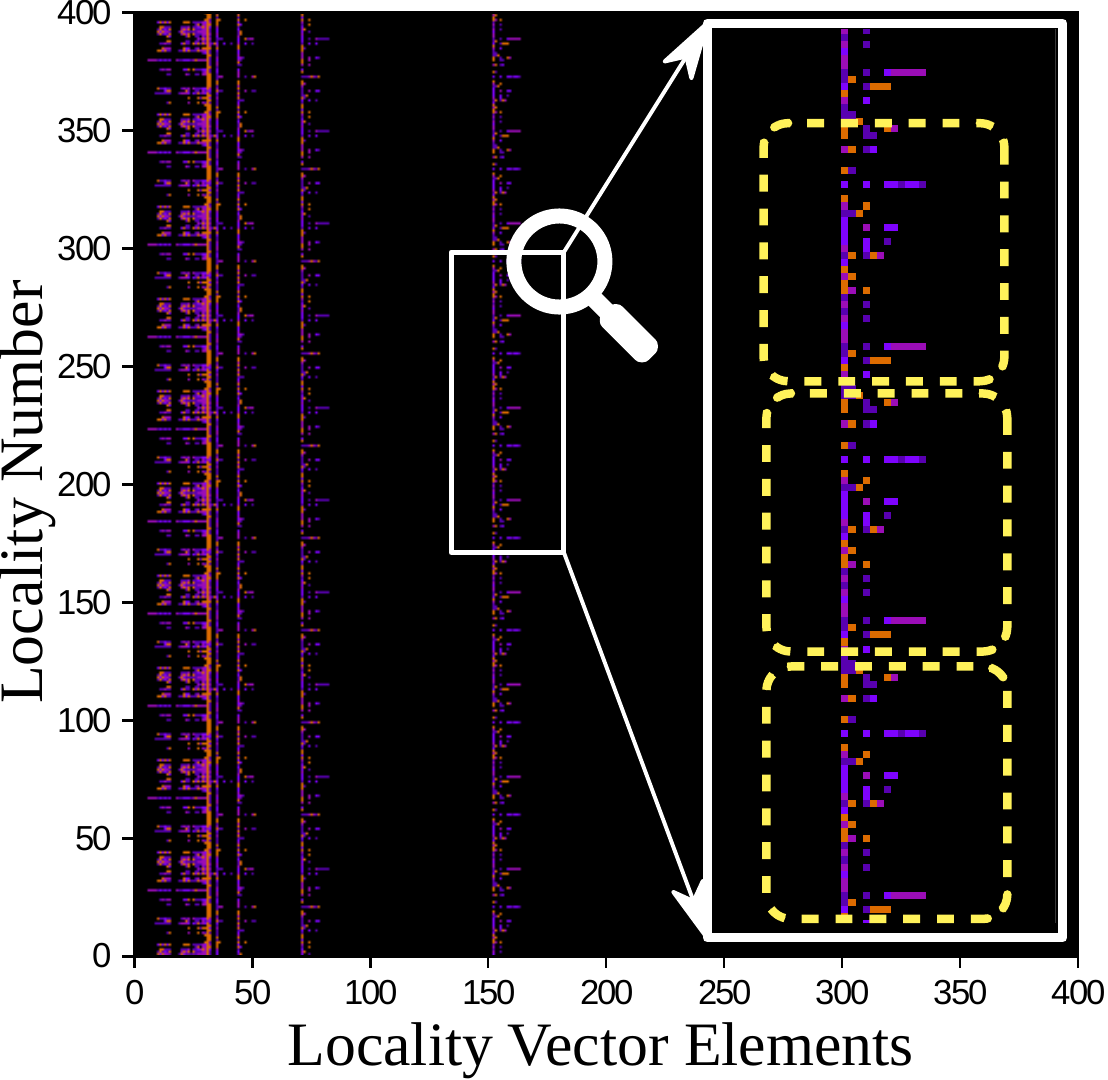}
	}
	\subfloat[1-Labeled Localities Image]{
		\includegraphics[width=0.35\columnwidth]{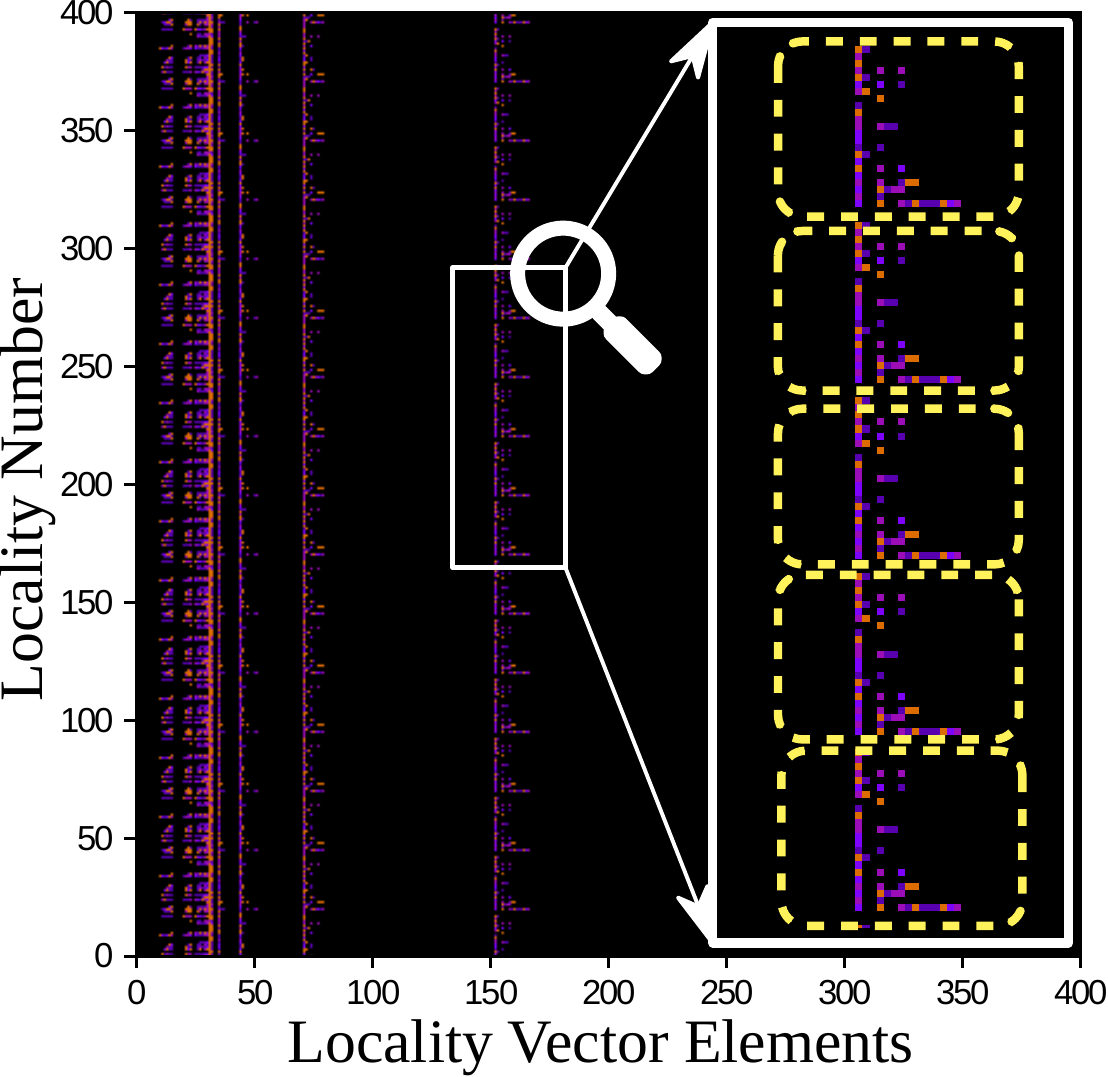}
	} 
	\caption{Vector Images for 0 and 1 Key Bit Values}
	\label{fig:patterns}
\end{figure}
\subsection{Application of ANNs} \label{CNNjustification}
To justify the selection of ANNs (in particular CNNs) for this particular problem, we present the following example. We perform the \textit{setup stage} of SnapShot for a randomly selected circuit locked with EPIC using 64-bit keys. In the next step, we extract 400 localities for both labels 0 and 1, using a pre-selected depth resulting in vectors of length 400. Finally, all 0 and 1 localities are stacked \textit{horizontally} to form two images: the first one represents all 0-labeled vectors, and the second represents all 1-labeled vectors. The result is shown in Fig.~\ref{fig:patterns}. Through careful examination, it is possible to determine \textit{repeating patterns} specific to each key bit value. Example patterns are marked in both images with dashed lines. In this case, the difference, i.e., the correlation between the key bit values and the represented images is clearly recognizable (even by a human). Since extracting and learning correlation in data is a key feature of CNNs, it stands to reason that CNNs are a suitable choice for this classification problem.

\section{Evaluation} \label{results}
\subsection{Experimental Setup} \label{experimental-setup}
\textbf{Benchmarks:} For evaluation, three open-source benchmark sets have been selected (Table~\ref{tab:benchmarks}): ISCAS'85~\cite{iscas2}, ITC'99~\cite{ITC99} and a set of modules selected from the 64-bit RISC-V processor core Ariane~\cite{ariane2019}. Note that ITC'99 contains sequential circuits with Flip Flops (FF). We evaluated both the generalized set scenario and the self-referencing scenario. SRS is relevant for comparing SnapShot to the state-of-the-art attack SAIL, while GSS has not been evaluated before. The GA is used for evolving CNNs for each benchmark and attack scenario \textit{separately}.
\begin{table}[b]\footnotesize
	\caption{Benchmark Circuits Used for Evaluation}
	\centering
	\subfloat[ISCAS'85 Benchmarks]{	\tabcolsep=0.12cm
		\begin{tabular}{cccc}\hline
			\textbf{IC} & \textbf{\#Ins} & \textbf{\#Gates} & \textbf{\#Outs} \\
			\hline
			\rowcolor{gray!20}	{c1355} & 41 & 546 & 32 \\
			{c1908} & 33 & 880 & 25 \\
			\rowcolor{gray!20}{c2670} & 233 & 1193 & 140 \\
			{c3540} & 50 & 1669 & 22 \\
			\rowcolor{gray!20}{c5315} & 178 & 2307 & 123 \\
			{c6288} & 32 & 2416 & 32 \\
			\rowcolor{gray!20}	{c7552} & 270 & 3512 & 108 \\ \hline
	\end{tabular}}\hfill\subfloat[ITC'99 Benchmarks]{	\tabcolsep=0.12cm
		\begin{tabular}{ccccc}\hline
			\textbf{IC} & \textbf{\#Ins} & \textbf{\#Gates} &\textbf{\#FFs} & \textbf{\#Outs} \\
			\hline
			\rowcolor{gray!20}	{b15} & 37 & 6931 & 447 & 70 \\
			{b21} & 34 & 7931 & 494 & 22 \\
			\rowcolor{gray!20}{b22} & 34 & 12128 & 709 & 22\\
			{b17} & 38 & 21191 & 1407 & 97 \\
			\rowcolor{gray!20}{b18} & 37 & 49293 & 3308 & 30\\
			{b19} & 47 & 98726 & 6618 & 40\\ \hline
	\end{tabular}}\hfill	\subfloat[Ariane RISC-V Modules]{	\tabcolsep=0.12cm
		\begin{tabular}{cccc}\hline
			\textbf{IC} & \textbf{\#Ins} & \textbf{\#Gates} & \textbf{\#Outs} \\
			\hline
			\rowcolor{gray!20}	{iscan} & 32 & 240 & 139 \\
			{commit} & 985 & 1584 & 417 \\
			\rowcolor{gray!20}	{brunit} & 342 & 1655 & 328 \\
			{decode}r & 518 & 2169 & 362 \\
			\rowcolor{gray!20}	{pcselect} & 521 & 3333 & 128 \\
			{brpredict} & 814 & 4669 & 333 \\
			\rowcolor{gray!20}	{alu} & 206 & 7412 & 65 \\\hline
	\end{tabular}}
	\label{tab:benchmarks}
\end{table}
\begin{table}[b]\footnotesize
	\caption{ANN Hyperparameters}
	\centering
	\subfloat[MLP]{	\tabcolsep=0.12cm
		\begin{tabular}{cc}\hline
			\textbf{Parameter} & \textbf{Value}  \\
			\hline
			\rowcolor{gray!20} {Layer Types} & Dense\\ 
			{Input Layer Act. Func.} & ReLU \\
			\rowcolor{gray!20} {Hidden Layer Act. Func.} & ReLU\\ 
			{Output Layer Act. Func.} & Softmax \\
			\rowcolor{gray!20}{GSS \#Units per Layer} & 400x1000x256x2  \\ 
			{SRS \#Units per Layer} & 400x512x128x2  \\
			\rowcolor{gray!20}{Batch Size} & 128  \\\hline
	\end{tabular}} \hfill \subfloat[CNN]{	\tabcolsep=0.12cm
		\begin{tabular}{cc}\hline
			\textbf{Parameter} & \textbf{Value} \\ \hline
			\rowcolor{gray!20}	Input/Conv./Dense Layer Act. Func. & ReLU \\
			Output Layer Act. Func. & Softmax \\
			\rowcolor{gray!20}Loss Functions & Sparse Cat. Cross Entropy \\
			Input/Conv. Layer Kernel Size & 3x3 \\
			\rowcolor{gray!20}\# Filters in Input/Conv. Layer & 64/128 \\
			\# Units in Dense/Output Layer & 128/2 \\
			\rowcolor{gray!20}	Pool/Stride Size & 2x2/1 \\
			Batch Size & 128 \\
			\rowcolor{gray!20}Optimizer/Learning Rate/Beta\_1/Beta\_2 & Adam/0.001/0.9/0.999 \\\hline
	\end{tabular}}
	\label{tab:cnn-hyperparameters}
\end{table}

\textbf{Environment:} The scripting is done using two Python frameworks: Keras/Tensorflow for deep learning and DEAP for evolutionary computation. The evaluation was performed on an Intel Core i5 CPU@3.2 GHz with 16 GB of RAM and a single Nvidia GeForce RTX 2080 Ti graphic card.

\textbf{Input Formatting:}~The localities are extracted using the described LVE procedure for the following parameters: $D_{b}=5$, $D_{f}=5$, $F_{in}=2$, $F_{out}=3$ and the encoding from Table~\ref{tab:encoding}. The extracted localities are trimmed to a length of 400 (removing excess empty locations) and normalized to values between 0 and 1. The MLP model uses these 1D vectors as input. To process the localities with the CNN model, the data is reshaped to the dimension $20x20x1$. The reshaping is done with the standard methods provided by the Python library numpy.

\textbf{Locking Scheme:}~The EPIC scheme has been selected to evaluate the effectiveness of SnapShot. The reasoning is twofold. First, EPIC is a \textit{superset} of other XOR/XNOR-based locking schemes, as other methods insert gates with a more \textit{selective} choice of locations compared to a random insertion, thereby inserting the \textit{same} key gates as EPIC. Therefore, the only differentiation to other schemes is the location selection. For example, SLL inserts key gates based on their relationship to hamper path sensitization attacks~\cite{sll1}. However, even if SLL performs a selective insertion, the key gates itself are identical to the ones in EPIC. Moreover, a random selection can inherently include all variations of location selections. The second reason is the generalization of the attack; a more selective insertion can only lead to more information leakage compared to a fully random choice.

\textbf{Performance Measure:} We evaluate the ANN prediction results using the key prediction accuracy (KPA) as defined in Section~\ref{snapshot:evolution}.

\subsection{MLP Model Setup}\label{MLPModelSetup}
As discussed earlier, MLPs typically consist of only a few hidden layers; otherwise the amount of trainable parameters easily explodes. We evaluated multiple MLP architectures manually by varying the number of hidden layers as well as the number of units per layer for both SRS and GSS. All layers are fully connected. The input layer consists of 400 units (same as locality length). The output layer consists of two outputs (one per key bit value). After the exploration, we fixed one architecture for each scenario based on the best performance achieved in the tuning phase. All MLP hyperparameters are listed in Table~\ref{tab:cnn-hyperparameters}~(a). In this case, due to no evident performance difference, the same MLP architecture is applied to all benchmarks in SRS. The final networks are trained for 100 epochs before deployment.

\subsection{CNN Evolution Setup} \label{cnn-setup}
\textbf{CNN Architecture:} The CNN evolution is performed as discussed in Section~\ref{snapshot:evolution}. All other fixed hyperparameters are listed in Table~\ref{tab:cnn-hyperparameters}~(b). We selected the standard parameters that are commonly used in literature for modern CNNs~\cite{cnn2017,cnn2015}. Dropout is not used since overfitting has not been observed in the evaluation. The maximum number of internal layers is limited to \textit{seven}. This value is selected through preliminary testing with up to fourteen layers. Moreover, the number of layers in the evolved CNNs (discussed in Section~\ref{res:networks}) tends to be much lower than seven, since a lower number of internal layers results in more efficient networks. This confirms that seven layers are sufficient for the architecture exploration for this particular problem.

\textbf{GA Setup:} All relevant GA parameters are listed in Table~\ref{tab:ga-params}. The GA parameters are selected based on a preliminary tuning phase as well as commonly used parameters and operators. However, both the notion and the implementation of all GA and CNN parameters are beyond the scope of this work and more details can be found in~\cite{LeCun1998, Eiben2015, neuroevolution2008}.

\textbf{Stopping Criterion:} In GA, every CNN architecture is trained for a selected number of epochs until the stopping criterion for training is reached. In order to tune the number of epochs to an amount that is \textit{sufficient for the CNN to converge} for all benchmarks, we trained a randomly selected architecture for each benchmark and attack scenario until an upper-limit of 80\% KPA is reached (Fig.~\ref{fig:convergence}). The reason why this limit has been selected is to extract the average amount of epochs necessary for all CNNs to converge to a reasonable KPA. Based on the results, 80\% upper-bound is sufficient to capture the learning asymptote and filter out fitter solutions. Consequently, the stopping criterion is set to 44 epochs. Note that this value is used in the training of CNNs during the GA exploration phase where it is not necessary to train the CNN instances to their limits. Once the best network is selected, the number of epochs can be increased before deployment.

\begin{table}[b]\footnotesize
	\begin{minipage}{0.45\linewidth}
		\caption{GA Parameters}
		\centering
		\tabcolsep=0.12cm
		\begin{tabular}{cc}\hline
			\textbf{Parameter} & \textbf{Value} \\ \hline
			GA Alg. Type & Generational \\
			\rowcolor{gray!20}Population Size & 10 \\
			Representation & Bitstring\\
			\rowcolor{gray!20}Generations & 20 \\
			Fitness & KPA \\
			\rowcolor{gray!20}Epochs & 44 \\
			Mutation Op. & Bit-Flip \\
			\rowcolor{gray!20}Mutation Prob. & 0.1 \\
			Crossover Prob. & 0.9 \\
			\rowcolor{gray!20}	Selection Op. & Tournament (3)\\
			Crossover Op. & Two-Point \\
			\rowcolor{gray!20} Gen. of Initial Population & Random \\\hline
		\end{tabular}
		\label{tab:ga-params}
	\end{minipage}\hfill
	\begin{minipage}{0.52\linewidth}
		\centering
		\begin{tikzpicture}[scale=0.88]
		\begin{axis}[
		ylabel style={at={(0.05,0.5)},anchor=north}, 
		legend image post style={scale=0.6},
		legend columns=2,
		ymajorgrids = true,
		width=7cm, height=6cm, 
		legend style={at={(0.49,0.38)},
			anchor=west,legend columns=-1},
		enlarge x limits={abs=0.6cm},
		ylabel style ={font=\large},
		xlabel style ={font=\large,at={(axis description cs:0.5,0.05)},anchor=north},	
		xtick align=inside,
		xmin=0,
		xmax=100,
		xtick={0, 20, 40, 60, 80, 100},
		ymin=50,ymax=85,
		ytick={50, 60,70, 80},	
		ylabel={KPA (\%)},
		xlabel={Epochs},
		major x tick style = {black, thin},
		major y tick style = {black, thin},
		minor tick length=1ex,
		]
		\draw[thick, dashed] (axis cs:\pgfkeysvalueof{/pgfplots/xmin},80) -- (axis cs:\pgfkeysvalueof{/pgfplots/xmax},80);
		\addplot[thick, dist1] plot coordinates { 
(0,57.341796)
(1,60.365236)
(2,61.04297)
(3,61.66992)
(4,61.955076)
(5,62.83594)
(6,63.400389999999994)
(7,63.640624)
(8,64.20116999999999)
(9,64.65625)
(10,65.384763)
(11,66.244143)
(12,66.88281)
(13,67.6875)
(14,68.73242)
(15,69.480467)
(16,69.92578)
(17,70.55078)
(18,71.11523)
(19,71.53711)
(20,71.91992400000001)
(21,72.38086)
(22,72.759765)
(23,73.01366999999999)
(24,73.67382599999999)
(25,73.77539)
(26,73.90625)
(27,74.458987)
(28,74.57422)
(29,74.80469)
(30,75.19336)
(31,75.28125)
(32,75.576174)
(33,75.97852)
(34,76.02539)
(35,76.45117)
(36,76.597655)
(37,76.55273700000001)
(38,76.857424)
(39,77.160156)
(40,77.316403)
(41,77.41405999999999)
(42,77.5625)
(43,77.64648)
(44,77.916014)
(45,78.14453)
(46,78.26366999999999)
(47,78.39258)
(48,78.287107)
(49,78.521484)
(50,78.55078)
(51,78.73047)
(52,78.791016)
(53,78.92383)
(54,79.089844)
(55,79.17187)
(56,79.26758000000001)
(57,79.271483)
(58,79.34961)
(59,79.472655)
(60,79.81055)
(61,79.53711)
(62,79.791015)
(63,79.914063)
(64,80.01953400000001)
};
		\addlegendentry{c1355}
		\addplot[thick, dist2] plot coordinates { 
(0,50)
(0,80)
};
		\addlegendentry{c1908}
		\addplot[thick, dist3] plot coordinates { 
(0,62.91992)
(1,67.25780999999999)
(2,69.384766)
(3,70.98633)
(4,72.277343)
(5,73.20312)
(6,74.20703)
(7,75.146484)
(8,75.63866999999999)
(9,76.51757599999999)
(10,77.14844)
(11,77.728516)
(12,78.0625)
(13,78.828126)
(14,79.308593)
(15,79.76758000000001)
(16,80.08008)
};
		\addlegendentry{c2670}
		\addplot[thick, dist4] plot coordinates { 
(0,56.96679999999999)
(1,65.64648)
(2,68.228513)
(3,69.890624)
(4,70.882815)
(5,72.21484000000001)
(6,73.01366999999999)
(7,74.02148)
(8,74.77929599999999)
(9,75.37109)
(10,76.13672)
(11,76.96289)
(12,77.71875)
(13,78.23828)
(14,78.70703300000001)
(15,79.2207)
(16,79.535156)
(17,80.228513)
};
		\addlegendentry{c3540}
		\addplot[thick, dist5] plot coordinates { 
(0,64.07226299999999)
(1,69.55078)
(2,71.910155)
(3,73.65820400000001)
(4,74.98828)
(5,75.84375)
(6,76.708984)
(7,77.38086)
(8,77.88086)
(9,78.42383000000001)
(10,78.902346)
(11,79.50195)
(12,80.17578)
};
		\addlegendentry{c5315}
		\addplot[thick, dist6] plot coordinates { 

(0,55.013674)
(1,72.02344000000001)
(2,79.947263)
(3,80)
};
		\addlegendentry{c6288}
		\addplot[thick, dist7] plot coordinates { 

(0,52.833985999999996)
(1,54.53125)
(2,56.251954999999995)
(3,57.507810000000006)
(4,58.64258)
(5,59.75)
(6,60.167970000000004)
(7,60.96289)
(8,61.69922)
(9,62.273437)
(10,62.607420000000005)
(11,63.023436)
(12,63.55274)
(13,64.064455)
(14,64.38672)
(15,65.17578400000001)
(16,65.296876)
(17,65.86132599999999)
(18,66.20312299999999)
(19,66.99414)
(20,67.31055)
(21,67.939454)
(22,68.226564)
(23,68.77148)
(24,69.10547)
(25,69.421875)
(26,69.70703)
(27,70.310545)
(28,70.65039)
(29,70.92578400000001)
(30,71.21875)
(31,71.65039)
(32,71.75781)
(33,72.21289)
(34,72.59179999999999)
(35,72.59961)
(36,72.77539)
(37,73.31836)
(38,73.519534)
(39,73.49804599999999)
(40,73.972654)
(41,73.95117)
(42,74.14844000000001)
(43,74.710935)
(44,74.742186)
(45,74.96679400000001)
(46,75.146484)
(47,75.210935)
(48,75.591797)
(49,75.64844)
(50,76.05077999999999)
(51,76.19141)
(52,76.125)
(53,76.5625)
(54,76.83008)
(55,76.84961)
(56,76.728517)
(57,77.04297)
(58,77.30078)
(59,77.14453300000001)
(60,77.43945)
(61,77.75195000000001)
(62,77.91211)
(63,77.80664)
(64,78.10352)
(65,78.30469)
(66,78.19922)
(67,78.27344000000001)
(68,78.509766)
(69,78.75195000000001)
(70,78.759766)
(71,78.933597)
(72,78.94922)
(73,79.16797)
(74,79.29492)
(75,79.408205)
(76,79.24023)
(77,79.60352)
(78,79.61328)
(79,79.88477)
(80,79.685545)
(81,79.95899)
(82,80.06445)
};
		\addlegendentry{c7552}
		\addplot[thick, dist8] plot coordinates { 
(0,52.30024456977844)
(1,58.0662727355957)
(2,64.38515186309814)
(3,69.59469318389893)
(4,73.25080037117004)
(5,75.74499845504761)
(6,78.09056043624878)
(7,79.58236932754517)
(8,80.93822598457336)
};

		\addlegendentry{b15}
		\addplot[thick, dist9] plot coordinates { 
(0,51.20311379432678)
(1,53.37532162666321)
(2,55.31664490699768)
(3,56.7946195602417)
(4,57.87365436553955)
(5,59.13802981376648)
(6,60.38826704025269)
(7,61.52070164680481)
(8,62.33115792274475)
(9,63.35836052894592)
(10,64.48137164115906)
(11,65.43161273002625)
(12,66.24677777290344)
(13,67.0807957649231)
(14,67.81114339828491)
(15,68.13312768936157)
(16,68.72369050979614)
(17,69.64095234870911)
(18,70.41370868682861)
(19,70.61474919319153)
(20,71.37965559959412)
(21,71.95765376091003)
(22,72.53565192222595)
(23,73.28171133995056)
(24,73.50788712501526)
(25,74.00735020637512)
(26,74.4706928730011)
(27,74.93717670440674)
(28,75.22460222244263)
(29,75.63297152519226)
(30,76.27222537994385)
(31,76.32719874382019)
(32,76.83608531951904)
(33,76.86750292778015)
(34,76.5957772731781)
(35,77.17534899711609)
(36,77.19733715057373)
(37,77.74392366409302)
(38,78.18998694419861)
(39,78.14286351203918)
(40,78.52767705917358)
(41,79.03342247009277)
(42,79.11666631698608)
(43,79.33341860771179)
(44,79.38681840896606)
(45,79.43236827850342)
(46,79.36325669288635)
(47,79.99936938285828)
(48,80.34648299217224)
};

		\addlegendentry{b21}
		\addplot[thick, col7] plot coordinates { 
(0,49.71911907196045)
(1,50.74847340583801)
(2,52.03922390937805)
(3,53.45986485481262)
(4,55.47636151313782)
(5,57.0496141910553)
(6,58.56280326843262)
(7,60.26431918144226)
(8,61.91875338554382)
(9,63.32640647888184)
(10,64.80386853218079)
(11,66.0069465637207)
(12,67.07202196121216)
(13,68.03805828094482)
(14,69.35965418815613)
(15,70.17470002174377)
(16,71.13423943519592)
(17,72.03857898712158)
(18,72.3600447177887)
(19,73.21730256080627)
(20,73.78555536270142)
(21,74.67690706253052)
(22,75.05682706832886)
(23,75.76470971107483)
(24,76.15761756896973)
(25,76.63170695304871)
(26,76.9271969795227)
(27,77.15449929237366)
(28,77.63020992279053)
(29,78.00688147544861)
(30,78.48584055900574)
(31,79.00052070617676)
(32,79.30737733840942)
(33,79.4340193271637)
(34,79.56552505493164)
(35,79.84153628349304)
(36,80.29289245605469)
};

		\addlegendentry{b22}
		\addplot[thick, dist11] plot coordinates { 
(0,51.56976580619812)
(1,54.90370392799377)
(2,57.58175849914551)
(3,60.97202301025391)
(4,63.74273300170898)
(5,66.41169786453247)
(6,68.6954915523529)
(7,70.50508856773376)
(8,72.30014801025391)
(9,73.51744174957275)
(10,74.60210919380188)
(11,75.69040656089783)
(12,76.72783136367798)
(13,77.32194662094116)
(14,78.21402549743652)
(15,78.85355949401855)
(16,79.45494055747986)
(17,80.01272082328796)
};

		\addlegendentry{b17}
		\addplot[thick, dist12] plot coordinates { 
(0,49.978014826774597)
(1,50.75370073318481)
(2,54.08333539962769)
(3,57.7401340007782)
(4,60.49284934997559)
(5,62.60004043579102)
(6,64.73009586334229)
(7,66.95688962936401)
(8,68.42558979988098)
(9,69.60582733154297)
(10,70.69987654685974)
(11,71.95398807525635)
(12,72.96712398529053)
(13,73.81492853164673)
(14,74.65569376945496)
(15,75.11125206947327)
(16,75.76205134391785)
(17,76.26510262489319)
(18,76.91942453384399)
(19,77.3749828338623)
(20,77.89386510848999)
(21,78.28786373138428)
(22,78.89469265937805)
(23,79.15677428245544)
(24,79.76184487342834)
(25,79.89376187324524)
(26,80.32293915748596)
};

		\addlegendentry{b18}
		\addplot[thick, dist13] plot coordinates { 
(0,52.46357321739197)
(1,56.48120641708374)
(2,59.08908247947693)
(3,61.15474104881287)
(4,62.75689601898193)
(5,64.05646204948425)
(6,65.35777449607849)
(7,66.63110256195068)
(8,67.53187775611877)
(9,68.5183584690094)
(10,69.27570700645447)
(11,70.14849781990051)
(12,70.68896293640137)
(13,71.70867323875427)
(14,72.55697846412659)
(15,73.3055830001831)
(16,74.0279495716095)
(17,74.50894713401794)
(18,75.36424398422241)
(19,76.00091099739075)
(20,76.39095187187195)
(21,76.82822346687317)
(22,77.50686407089233)
(23,78.04558277130127)
(24,78.6017894744873)
(25,79.10552024841309)
(26,79.500812292099)
(27,80.11823892593384)
};

		\addlegendentry{b19}		
		\end{axis}
		\end{tikzpicture}
		
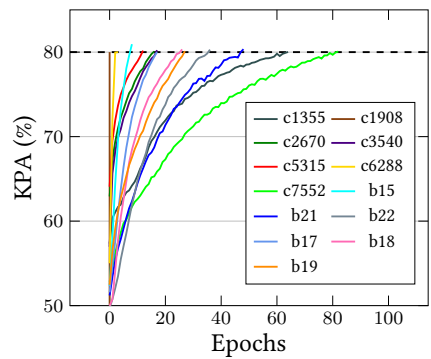
\captionof{figure}{Convergence Graph}
		\label{fig:convergence}
	\end{minipage}
\end{table}
\subsection{Results: Generalized Set Scenario}
For GSS, we trained the constructed MLP and the evolved CNNs using a training set compiled of all ISCAS'85 and ITC'99 benchmarks mentioned in Table~\ref{tab:benchmarks}. Each training benchmark is locking 1,000 times with 64-bit keys, resulting in a total of 832,000 training localities. The test set consists of all selected Ariane modules (Table~\ref{tab:benchmarks}~(c)), each locked 1,000 times with random 64-bit keys; resulting in a total of 448,000 test localities (64,000 per target). 

\begin{figure}
	\centering
	\subfloat[GSS Evaluation]{
		\begin{tikzpicture}[scale=0.8]
		\pgfplotstableread[row sep=\\,col sep=&]{			
			interval & MLP-SnapShot & CNN-SnapShot \\
			decoder & 55.77  & 58.64  \\
			pcselect & 72.37 & 77.55  \\ 
			brpredict  & 57.83 & 65.21  \\
			alu  & 59.34 & 63.30  \\
			iscan & 50.11 & 51.78 \\
			commit & 57.51 & 61.19 \\
			brunit & 51.09 & 53.27\\
		}\mydata
		\begin{axis}[
		ylabel style={at={(0.025,0.5)},anchor=north}, 
		legend image post style={scale=0.8},
		ybar,	legend columns=3,
		ymajorgrids = true,
		bar width=.4cm,
		width=12cm, height=5cm, 
		legend style={at={(0.27,0.98)},
			anchor=north,legend columns=-1},
		symbolic x coords={iscan, commit, brunit, decoder, pcselect, brpredict, alu},
		enlarge x limits={abs=0.6cm},
		ylabel style ={font=\large},
		xlabel style ={font=\large},
		every node near coord/.append style={
			anchor=north,
			yshift=2.5ex,
			xshift=-1.5ex,
			font=\footnotesize,
			rotate=90
		},
		x tick style={draw=none},
		every node near coord/.append style={
			/pgf/number format/fixed, 
			/pgf/number format/fixed zerofill,
			/pgf/number format/precision=2
		},
		ytick={50, 60, 70, 80, 90, 100},	
		xtick align=inside,
		xtick=data,
		ymin=50,ymax=110,
		ylabel={KPA (\%)},
		xlabel={RISC-V Ariane Modules},
		major x tick style = {black, thin},
		major y tick style = {black, thin},
		minor tick length=1ex,
		]
		\addplot[draw=black,fill=white, nodes near coords] table[x=interval,y=MLP-SnapShot]{\mydata};
		\addplot[draw=black,fill=white, nodes near coords, postaction={pattern=north east lines}] table[x=interval,y=CNN-SnapShot]{\mydata};
		\legend{MLP-SnapShot, CNN-SnapShot}
		\end{axis}
		\end{tikzpicture}
	} \qquad	\subfloat[Average KPA (GSS)]{
		\begin{tikzpicture}[scale=0.8]
		\begin{axis}[
		ylabel style={at={(0.1,0.5)},anchor=north}, 
		legend image post style={scale=0.8},
		ybar,	legend columns=2,
		ymajorgrids = true,
		bar width=.4cm,
		width=4cm, height=5cm, 
		legend style={at={(0.36,0.98)},
			anchor=north},
		ylabel style ={font=\large},
		xlabel style ={font=\large},
		every node near coord/.append style={
			anchor=north,
			yshift=2.5ex,
			xshift=-1.5ex,
			font=\footnotesize,
			rotate=90
		},
		x tick style={draw=none},
		every node near coord/.append style={
			/pgf/number format/fixed, 
			/pgf/number format/fixed zerofill,
			/pgf/number format/precision=2
		},
		xmin=0,
		xmax=4,
		xtick={1,3},
		x tick style={draw=none},
		xticklabels={MLP, CNN},
		ytick={50, 60, 70, 80, 90, 100},	
		ymin=50,ymax=110,
		ylabel={YYY},
		ylabel={KPA (\%)},
		xlabel={Attack},
		major x tick style = {black, thin},
		major y tick style = {black, thin},
		minor tick length=1ex,
		every axis plot/.append style={
			ybar,
			bar shift=0pt,
			fill
		}]
		
		\addplot[draw=black,fill=white, nodes near coords, forget plot]coordinates {(1,57.71)};
		
		
		\addplot[draw=black,fill=white, nodes near coords, postaction={pattern=north east lines}]coordinates {(3,61.56)};
		
		\legend{}
		\end{axis}
		\end{tikzpicture}
	}
	\vspace{-0.2cm}
	\subfloat[SRS Evaluation]{
		\begin{tikzpicture}[scale=0.8]
		\pgfplotstableread[row sep=\\,col sep=&]{			
			interval & MLP-SnapShot & SAIL & CNN-SnapShot \\
			c1355 & 54.53 & 55.83 & 78.80  \\
			c1908 & 57.73 & 63.33 &  79.02 \\ 
			c2670 & 68.59 & 75.83  & 78.43 \\
			c3540 & 74.21 & 72.08  & 85.94 \\
			c5315 & 68.82 & 74.79  & 79.92 \\
			c6288 & 88.90 & 93.75 & 96.87 \\
			c7552 & 63.28 & 69.16  &  79.25 \\
		}\mydata
		\begin{axis}[
		ylabel style={at={(0.018,0.5)},anchor=north}, 
		legend image post style={scale=0.8},
		ybar, legend columns=3,
		ymajorgrids = true,
		bar width=.4cm,
		width=14cm, height=5cm, 
		legend style={at={(0.27,0.98)},
			anchor=north,legend columns=-1},
		symbolic x coords={c1355, c1908, c2670, c3540, c5315, c6288, c7552, b15, b17, b18, b19, b21, b22},
		enlarge x limits={abs=0.8cm},
		ylabel style ={font=\large},
		xlabel style ={font=\large},
		every node near coord/.append style={
			anchor=north,
			yshift=2.5ex,
			xshift=-1.5ex,
			font=\footnotesize,
			rotate=90
		},
		x tick style={draw=none},
		every node near coord/.append style={
			/pgf/number format/fixed, 
			/pgf/number format/fixed zerofill,
			/pgf/number format/precision=2
		},
		ytick={50, 60, 70, 80, 90, 100},	
		xtick align=inside,
		xtick=data,
		ymin=50,ymax=110,
		ylabel={KPA (\%)},
		xlabel={ISCAS'85 Benchmarks},
		major x tick style = {black, thin},
		major y tick style = {black, thin},
		minor tick length=1ex,
		]
		
		\addplot[draw=black,fill=white, nodes near coords] table[x=interval,y=MLP-SnapShot]{\mydata};
		\addplot[draw=black,fill=gray!50, nodes near coords] table[x=interval,y=SAIL]{\mydata};
		\addplot[draw=black,fill=white, nodes near coords, postaction={pattern=north east lines}] table[x=interval,y=CNN-SnapShot]{\mydata};
		\legend{MLP-SnapShot, SAIL, CNN-SnapShot}
		\end{axis}
		\end{tikzpicture}
	}	\subfloat[Average KPA (SRS)]{
		\begin{tikzpicture}[scale=0.8]
		\begin{axis}[
		ylabel style={at={(0.07,0.5)},anchor=north}, 
		legend image post style={scale=0.8},
		ybar,	legend columns=2,
		ymajorgrids = true,
		bar width=.4cm,
		width=4cm, height=5cm, 
		legend style={at={(0.36,0.98)},
			anchor=north},
		ylabel style ={font=\large},
		xlabel style ={font=\large},
		enlarge x limits={abs=0.1cm},
		every node near coord/.append style={
			anchor=north,
			yshift=2.5ex,
			xshift=-1.5ex,
			font=\footnotesize,
			rotate=90
		},
		x tick style={draw=none},
		every node near coord/.append style={
			/pgf/number format/fixed, 
			/pgf/number format/fixed zerofill,
			/pgf/number format/precision=2
		},
		xmin=0,
		xmax=4,
		xtick={0.5,2,3.5},
		x tick style={draw=none},
		xticklabels={MLP, SAIL, CNN},
		ytick={50, 60, 70, 80, 90, 100},	
		ymin=50,ymax=110,
		ylabel={YYY},
		ylabel={KPA (\%)},
		xlabel={Attack},
		major x tick style = {black, thin},
		major y tick style = {black, thin},
		minor tick length=1ex,
		every axis plot/.append style={
			ybar,
			bar shift=0pt,
			fill
		}]
		
		\addplot[draw=black,fill=white, nodes near coords, forget plot]coordinates {(0.5,68.00)};
		
		\addplot[draw=black,fill=gray!50, nodes near coords]coordinates {(2,72.11)};
		
		\addplot[draw=black,fill=white, nodes near coords, postaction={pattern=north east lines}]coordinates {(3.5,82.60)};
		
		\legend{}
		\end{axis}
		\end{tikzpicture}
	}
	
	\vspace{-0.2cm}
	\subfloat[SRS Evaluation]{
		\begin{tikzpicture}[scale=0.8]
		\pgfplotstableread[row sep=\\,col sep=&]{			
			interval & MLP-SnapShot & CNN-SnapShot \\
			b15   & 78.06  & 80.59 \\
			b21   & 76.68  & 80.37 \\
			b22   & 76.96  & 81.34 \\
			b17   & 71.64  & 79.92 \\
			b18   & 75.55 & 81.11 \\
			b19   & 75.57  & 80.14 \\
		}\mydata
		\begin{axis}[
		ylabel style={at={(0.03,0.5)},anchor=north}, 
		legend image post style={scale=0.8},
		ybar, legend columns=3,
		ymajorgrids = true,
		bar width=.4cm,
		width=10cm, height=5cm, 
		legend style={at={(0.33,0.98)},
			anchor=north,legend columns=-1},
		symbolic x coords={c1355, c1908, c2670, c3540, c5315, c6288, c7552, b15, b21, b22, b17, b18, b19},
		enlarge x limits={abs=0.6cm},
		ylabel style ={font=\large},
		xlabel style ={font=\large},
		every node near coord/.append style={
			anchor=north,
			yshift=2.5ex,
			xshift=-1.5ex,
			font=\footnotesize,
			rotate=90
		},
		x tick style={draw=none},
		every node near coord/.append style={
			/pgf/number format/fixed, 
			/pgf/number format/fixed zerofill,
			/pgf/number format/precision=2
		},
		ytick={50, 60, 70, 80, 90, 100},	
		xtick align=inside,
		xtick=data,
		ymin=50,ymax=110,
		ylabel={KPA (\%)},
		xlabel={ITC'99 Benchmarks},
		major x tick style = {black, thin},
		major y tick style = {black, thin},
		minor tick length=1ex,
		]
		
		\addplot[draw=black,fill=white, nodes near coords] table[x=interval,y=MLP-SnapShot]{\mydata};
		\addplot[draw=black,fill=white, nodes near coords, postaction={pattern=north east lines}] table[x=interval,y=CNN-SnapShot]{\mydata};
		\legend{MLP-SnapShot, CNN-SnapShot}
		\end{axis}
		\end{tikzpicture}
	}\qquad	\subfloat[Average KPA (Total SRS)]{
		\hspace*{4pt}%
		\begin{tikzpicture}[scale=0.8]
		\begin{axis}[
		ylabel style={at={(0.1,0.5)},anchor=north}, 
		legend image post style={scale=0.8},
		ybar,	legend columns=2,
		ymajorgrids = true,
		bar width=.4cm,
		width=4cm, height=5cm, 
		legend style={at={(0.36,0.98)},
			anchor=north},
		ylabel style ={font=\large},
		xlabel style ={font=\large},
		every node near coord/.append style={
			anchor=north,
			yshift=2.5ex,
			xshift=-1.5ex,
			font=\footnotesize,
			rotate=90
		},
		x tick style={draw=none},
		every node near coord/.append style={
			/pgf/number format/fixed, 
			/pgf/number format/fixed zerofill,
			/pgf/number format/precision=2
		},
		xmin=0,
		xmax=4,
		xtick={1,3},
		x tick style={draw=none},
		xticklabels={MLP, CNN},
		ytick={50, 60, 70, 80, 90, 100},	
		ymin=50,ymax=110,
		ylabel={YYY},
		ylabel={KPA (\%)},
		xlabel={Attack},
		major x tick style = {black, thin},
		major y tick style = {black, thin},
		minor tick length=1ex,
		every axis plot/.append style={
			ybar,
			bar shift=0pt,
			fill
		}]
		
		\addplot[draw=black,fill=white, nodes near coords, forget plot]coordinates {(1,71.57)};
		
		
		\addplot[draw=black,fill=white, nodes near coords, postaction={pattern=north east lines}]coordinates {(3,81.67)};
		
		\legend{}
		\end{axis}
		
		\end{tikzpicture}
		\hspace*{10pt}%
	}\hspace{1cm}
	\caption{Attack Evaluation for the Generalized Set and Self-Referencing Scenario}
	\label{fig:eval}
\end{figure}
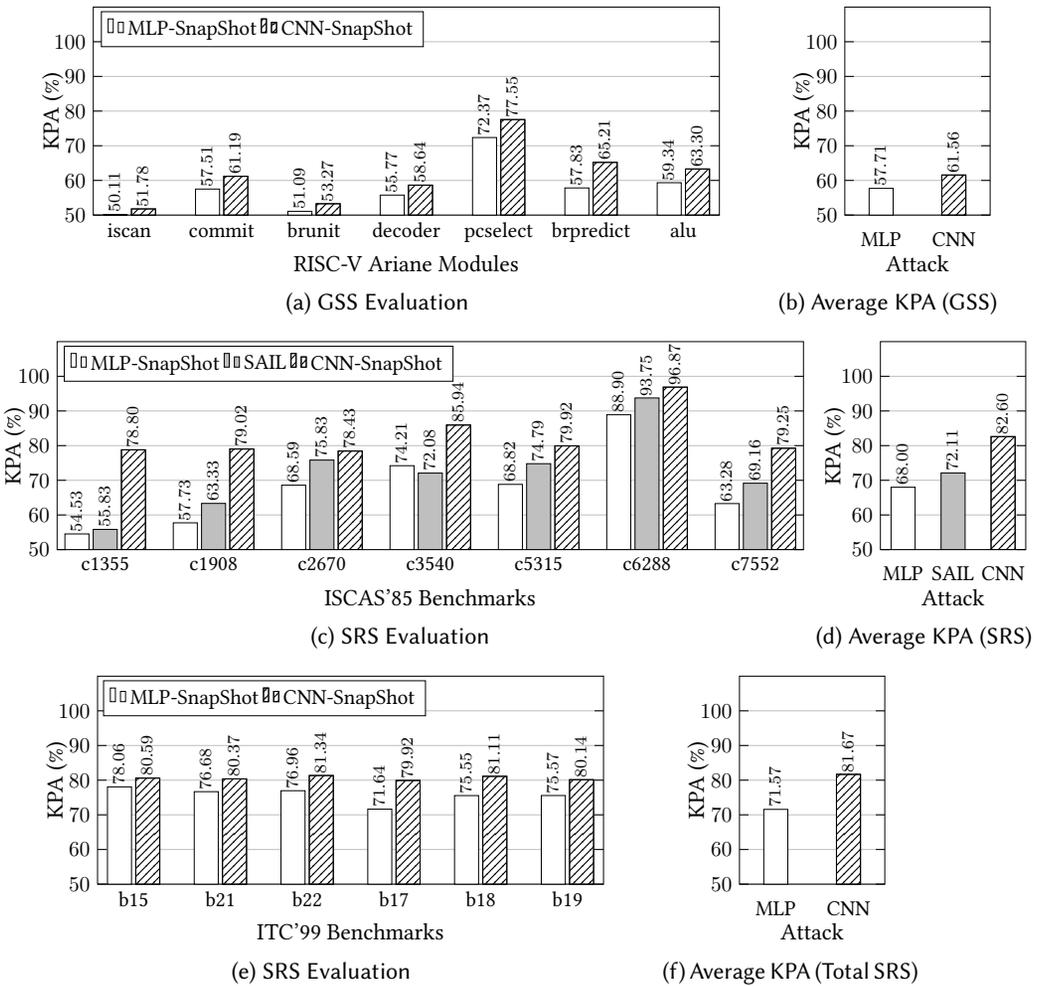

The average test KPA (Fig.~\ref{fig:eval}~(b)) across all modules for GSS is 57.71\% for MLP, and 61.56\% for CNN, with a maximum of 77.55\% for the $pcselect$ module (Fig.~\ref{fig:eval}~(a)). This indicates the feasibility of learning from other locked netlists to predict a key of a previously unseen netlist with an average increase in accuracy of up to 11.56 percentage points compared to a random guess. Compared to MLP, the CNN model offers a relatively small performance gain of 3.85\% percentage points on average. In general, GSS offers one advantage over SRS; only one universal training set needs to be assembled regardless of the target, i.e, the relocking of the target netlist is avoided.

\subsection{Results: Self-Referencing Scenario}
SRS describes a more powerful scenario; learning and predicting on the same locked netlist. We locked all ISCAS'85 and ITC'99 benchmarks shown in Table~\ref{tab:benchmarks} with 64-bit (target) keys. The training set is generated by \textit{relocking} the already locked benchmarks repeatedly 1,000 times with 64-bit keys, resulting in 64,000 training localities per benchmark and iteration instance.

The achieved average test KPA for the SnapShot MLP and CNN models is shown in Fig.~\ref{fig:eval}~(c) in comparison to the state-of-art SAIL attack. The average KPA of CNN-SnapShot is 82.60\%, resulting in a significant improvement of 10.49 percentage points compared to SAIL, as shown in Fig.~\ref{fig:eval}~(d). Note that this average is based only the ISCAS'85 benchmarks to be consistent with the SAIL evaluation. Moreover, SnapShot outperforms SAIL for \textit{all} evaluated benchmarks. Another important observation is that SnapShot achieves a minimum KPA of 78.43\% (for $c2670$), whereas SAIL achieves a KPA of 55.83\% (for $c1355$); only 5.83 percentage points more than a random guess. Therefore, the evolved CNNs are able to achieve a high KPA regardless of the circuit type. However, circuits with more regular structures, such as $c6288$, result in a higher KPA since additional logic is easier detected in repeating structures.

The MLP-based approach in SnapShot achieves a relatively low average KPA of 68.00\% (Fig.~\ref{fig:eval}~(d)), with lower performance values for almost all benchmarks compared to SAIL, and a lower performance compared to all CNN-based KPAs. 

To demonstrate the scalability of SnapShot, in addition to the values in (Fig.~\ref{fig:eval}~(c)), we performed another evaluation of SnapShot-based MLP and CNN attacks using larger sequential benchmarks. As shown in Fig.~\ref{fig:eval}~(e), both SnapShot variants result in consistently high KPA values (above 71.64\% for MLP and 79.92\% for CNN). However, similarly to the previous evaluation, the evolved CNNs outperform the MLP model for all benchmarks. The total average KPA across \textit{all} combinational (ISCAS'85) and sequential (ITC'99) circuits using SnapShot is 71.57\% for MLP and 81.67\% for the CNNs (Fig.~\ref{fig:eval}~(d)). Interestingly, the CNN approach achieves similar performance values regardless of the circuit type. \textit{This strongly suggests that convolutional neural networks offer a reliable tool for attacking XOR/XNOR-based logic locking schemes, while using a simpler and more flexible prediction model compared to the state of the art.} Moreover, the evaluation indicates that CNNs offer consistently better performance values compared to standard MLPs for this particular key prediction problem.

Furthermore, compared to GSS, SnapShot achieves significantly higher KPA values for SRS. This comparison offers novel insights into a previously untested scenario. Thereby, we can conclude that SRS enables a more focused learning process targeted towards the specific locked netlist irrespective of the underlying ML-model, \textit{offering a more favorable attack scenario for an adversary}. 

\subsection{Evolved CNN Architectures}\label{res:networks}
The best evolved architectures for all benchmarks and scenarios are presented in Table~\ref{tab:GSS-best-solutions} and Table~\ref{tab:SRS-best-solutions} in the form of genotypes. For GSS, only one final network is evolved since the training set is equivalent for all benchmarks. In comparison, SRS yields a single network for each benchmarks. Interestingly, the evolved networks have a maximum of only five internal layers for GSS and up to three layers for SRS. This offers the following insights. First, the GSS training set is more difficult to learn, i.e., it requires a more refined feature extraction. This corresponds to the way the training sets are generated, as in GSS a single large set of different relocked benchmarks is used while in SRS a self-referencing set is generated. Secondly, the evolutionary process suggests that using a smaller number of layers is sufficient, as no genotype using all layers has been reported as a fitter solution. Therefore, starting with a limit of seven layers is more than enough for this particular problem. Furthermore, the evaluated stopping criterion of only 44 epochs (Section~\ref{cnn-setup}) indicates that the training process easily converges. These observations suggest that the classification of XOR and XNOR localities to their correct activation bits is not a particularly difficult problem. 
\textit{Consequently, the results imply that the \textbf{security foundation} of XOR/XNOR-based locking schemes is questionable and that it must be revised.} 

All networks can be reproduced based on the provided genotypes, hyperparameters (Table~\ref{tab:cnn-hyperparameters}), fixed layers (Section~\ref{cnn-setup}) as well as open-source benchmarks (Section~\ref{experimental-setup}).
\begin{table}[!t]\footnotesize
	\caption{Evolved CNN Architecture (Genotype) for GSS}
	\centering
	\begin{tabular}{cc}\hline
		\textbf{IC} & \textbf{[[Layer Presence], [Layer Type]]} \\
		\hline
		\rowcolor{gray!20}Training Benchmarks & [[0, 1, 1, 1, 1, 0, 1], [0, 1, 1, 1, 1, 1, 0]]  \\\hline
	\end{tabular}
	\label{tab:GSS-best-solutions}
\end{table}

\begin{table}[t]\footnotesize
	\caption{Evolved CNN Architectures (Genotypes) for SRS}
	\centering
	\subfloat[ISCASC'85]{	\tabcolsep=0.12cm
	\centering
	\begin{tabular}{cc}\hline
		\textbf{IC} & \textbf{[[Layer Presence], [Layer Type]]} \\
		\hline
		\rowcolor{gray!20}c1355 & [[0, 1, 0, 0, 0, 0, 1], [1, 0, 1, 1, 1, 0, 1]] \\
		c1908 & [[0, 0, 0, 1, 0, 0, 0], [0, 0, 1, 1, 0, 0, 0]] \\
		\rowcolor{gray!20}c2670 & [[0, 1, 0, 0, 0, 0, 0], [0, 0, 1, 1, 1, 1, 1]] \\
		c3540 & [[0, 0, 0, 0, 1, 1, 0], [1, 0, 0, 0, 1, 0, 0]] \\
		\rowcolor{gray!20}c5315 & [[0, 0, 0, 0, 1, 1, 0], [1, 0, 0, 0, 1, 0, 0]]\\
		c6288 & [[0, 0, 0, 0, 1, 1, 0], [1, 0, 0, 0, 1, 0, 0]]\\
		\rowcolor{gray!20}c7552 & [[1, 1, 1, 0, 0, 0, 0], [0, 1, 0, 1, 0, 0, 1]]\\\hline
\end{tabular}}\hfil \subfloat[ITC'99]{	\tabcolsep=0.12cm
\centering
\begin{tabular}{cc}\hline
	\textbf{IC} & \textbf{[[Layer Presence], [Layer Type]]} \\
	\hline
	\rowcolor{gray!20}	b15 & [[0, 0, 0, 0, 0, 0, 1], [1, 0, 0, 0, 1, 1, 1]]\\
b21 & [[0, 1, 0, 0, 0, 0, 0], [1, 1, 1, 0, 1, 0, 0]]\\
	\rowcolor{gray!20}b22 & [[0, 0, 0, 0, 0, 0, 1,], [1, 1, 1, 0, 0, 0, 1]]\\
	b17 & [[0, 0, 1, 0, 0, 0, 0], [1, 1, 1, 0, 0, 0, 0]]\\
	\rowcolor{gray!20}b18 & [[0, 0, 0, 0, 0, 0, 1], [0, 0, 0, 0, 0, 1, 1]]\\
	b19 & [[0, 0, 1, 0, 0, 0, 0], [1, 0, 1, 0, 0, 1, 1]]\\\hline
\end{tabular}
	}\label{tab:SRS-best-solutions}
\end{table}
\subsection{Training Time Analysis}
The neuroevolutionary mechanism facilitates the process of defining suitable layer configurations for the feature extraction component of the neural network. Using the neuroevolutionary process, an adversary is able to search through only a fraction of the possible configurations and still achieve valuable results. This mirrors the scenario in which an adversary has only a limited amount of time to acquire the correct activation key and proceed with reverse engineering. For example, after the design has been sent to the foundry, typically only a few months are needed until the fabrication is done. This window greatly limits the  available time frame for performing an attack, since a malicious change still has to be implemented before the final fabrication. Therefore, having an attack that can converge more quickly is crucial for the success of the adversary.

\textbf{Genotype and Phenotype Search Space:} The limit of 7 internal layers yields a genotype search space of $2^{14}$ different configurations, since 7 bits are needed to represent the existence of each layer and 7 bits to represent their types. However, some different genotypes can map to an equivalent phenotype. This is the case if a particular layer presence bit $x_{i}^{p}=0$. In this case, the corresponding layer type $x_{L+i}^{t}$ is not relevant for the construction of the phenotype (CNN architecture). Therefore, the total phenotype search space is $3^{7} = 2,187$. However, during the mutation and crossover procedures, even different genotypes with an equivalent phenotype store potentially valuable (and different) genetic material. Therefore, these genotypes must still be considered in the genetic process. Nonetheless, to avid a dependency on the genotype, in the following we discuss the timing comparison in terms of the phenotype search space. 

\textbf{Iterative Search:} As mentioned, an iterative search has to consider $3^{7}$ different configurations.

\textbf{Neuroevolution:} The size of the search space using the neuroevolutionary mechanism is defined by two GA parameters (Table~\ref{tab:ga-params}): the number of generations and the size of the population. Since a generational GA is selected, in every generation, a new population is generated. This amounts to $20\cdot10=200$ configurations, i.e., only 9.14\% of the total search space.

\textbf{Time Comparison:} The evaluation of a \textit{single} configuration takes 44 epochs. Assuming the existence of a single GPU (see Section~\ref{experimental-setup}), one epoch takes approximately 10 seconds. An iterative search using this setup amounts to 11.13 days. In comparison, the neuroevolution takes 1.01 days. In practice, neuroevolution takes even less time since equivalent genotypes are only evaluated once and the fitness function can be adapted to detect phenotypes that have been evaluated before. It is evidently possible to distribute the search using multiple GPUs. However, this has not been evaluated in this work as using a single GPU offers clear timing insights into the attack.

\section{Discussion and Analysis}\label{discussion}
To design learning-resilient logic locking, understanding the reasons for the vulnerabilities induced by existing locking schemes is essential. Let us look at two aspects of logic locking: the location selection and the introduced change. The location selection indicates how a locking scheme \textit{decides} where a change is introduced. If the selection or the actual change exposes information in some form, an entity (human or artificial system) is able to learn about the key. Therefore, a locking scheme that has no biased location selection and introduces unpredictable changes is learning-resilient. To test this claim, we performed the following experiment. We locked multiple benchmarks with a modified version of EPIC with randomly generated 64-bit keys, without another synthesis round. The modified scheme inserts XOR/XNOR gates \textit{regardless} of the key bit value. Obviously, this scheme is functionally incorrect, however it serves its purpose; SnapShot reports KPA values of 50\% for all benchmarks. In other words, the selected locking scheme forces the CNN to randomly guess the key, as no information is leaked through the location selection or the introduced change, as both are completely random and unpredictable. 

From a different perspective, this experiment shows that a locking scheme can be learning-resilient if \textit{two identical localities} (same vector image) for two different key inputs result in \textit{different labels}, i.e., two equal images suggesting different labels. Consequently, such a behavior enforces a random guess. \textit{Based on this analysis, we can conclude that a learning-resilient locking scheme should be resilient against guessing attacks even before resynthesis, i.e., it should not depend on the changes induced by a synthesis tool}.

Interestingly, the experiment also suggests that EPIC leaks information based on the introduced change, not the location selection. However, for EPIC, the information leakage is straightforward. XOR is inserted for a key bit 0, and XNOR (XOR + inverter) for a key bit 1. This scheme implies that an XOR locality is predicted to be a 1 only if XOR is by chance placed as input to an inverter (similar for XNOR and 0). However, this case does not appear very often.

Moreover, a functionally correct learning-resilient scheme is revealed when looking at the cases where SnapShot made a false prediction. For example, when two key gates are inserted (by chance) back to back, they form a run of gates, e.g., XOR-XOR for the key 00. In this case, SnapShot is not able to guess whether the key is 00 or 11 as both result in a correct buffering. However, a run of gates can easily be detected and replaced by a single gate, thereby dissolving the learning-resilience.

Another conclusion can be drawn based on the attack complexity. SnapShot (as well as SAIL) can attack each key input individually since the key gates typically exhibit a very local dependency without interacting with other key-driven gates in the netlist. The reason for this is simple; each key gate is driven by a separate key input which does not rely on other key inputs for its correctness. Increasing the complexity of the mutual dependencies among key inputs also increases the difficulty of guessing a correct key bit, since (in theory) multiple key bits must be considered at the same time. A possible mitigation can be found in embedded key-generation units that ensure a highly interdependent key set~\cite{enhancing2017}. However, these units represent an add-on mitigation. Therefore, it is more relevant to look into schemes that are learning-resilient by design. The learning resiliency of key-generation units remains to be evaluated.

\textbf{Lessons Learned:~} Based on the provided data and analysis, we can conclude that ML-based attacks offer a novel pathway to uncover critical vulnerabilities in logic locking schemes. The SnapShot attack has showcased that it is possible to successfully attack the fundamental security assumption of XOR-based locking, leading to the conclusion that an adversary can make an educated guess about the key bits based on the inserted key gate types. Evidently, this goes against the assumed security foundation of XOR-based locking. Interestingly, a vast amount of attacks, including the most powerful SAT-based attacks~\cite{decadeOfLocking}, do not rely on this assumption at all, but rather focus on secondary security aspects of locking schemes (e.g., functional corruption for incorrect keys). Challenging the security foundation with machine learning can lead to novel insights for the design of next-generation locking schemes and a new landscape of key-recovery attacks.

\section{Limitations and Research Opportunities}\label{futurework}
A limitation of prediction-based attacks is the uncertainty of the prediction for each key bit. Even though the key can be retrieved with a high accuracy, it still remains a challenge to determine \textit{which key bits} are correctly predicted. However, the predicted result can be used as a seed for subsequent attacks to further refine the key, as showcased in~\cite{SURF2019}. Furthermore, a predicted key can be fixed in the design to "remove" the redundant logic with reoptimization. This step can facilitate the process of identifying different components of a larger design through similarity matching (an important step in modern reverse engineering~\cite{MLReverseEngineering}).

The attack flow of SnapShot includes a black box optimization mechanism. Therefore, the flow can be extended for automatic attack exploration using a variety of representations (locality extraction strategies) and ML models for supervised learning. However, due to the nature of the representation (Section~\ref{snapshot:extraction}) as well as the existing correlation in data (Section~\ref{CNNjustification}), neural networks are a suitable tool for this problem.

Furthermore, GSS outlines a new research opportunity; assembling a generic benchmark suite covering a large spectrum of locked netlists. In the long run, this benchmark suite can be utilized for training a model to automatically design learning-resilient schemes. Moreover, an interesting research task is to analyze the impact of the locality size on the prediction accuracy.

Recently, a great variety of more advanced logic locking schemes has been introduced that offer protection even against SAT-based attacks~\cite{fullLock2019,strongAntiSat2020, evo2017}. However, some of these schemes, such as Anti-SAT~\cite{antisat2017} and CAS-Lock~\cite{caslock2019}, rely on the mentioned XOR/XNOR property to encode the correct user-defined key. Therefore, the evaluation in this work provides insights into a fundamental security issue that even spans across the latest logic locking schemes. Nonetheless, directly attacking these schemes with SnapShot remains part of the future work.

Another potentially suitable deep learning model for attacking logic locking schemes are Graph Neural Networks (GNNs)~\cite{gnns2018}. These networks perform inference on data that is described in the form of graphs. GNNs have been evaluated on  circuit-related problems as well~\cite{GNNDistributedCircuitDesign2019}. In particular, GNNs have been used for attack time estimation on logic locking~\cite{EstimatingDeObfWithGNN2020}. In this sense, GNNs offer a potentially fruitful research ground in the domain of hardware security. Moreover, it might be possible to transfer existing convolutional neural architectures to this specific classification problem. However, since this problem domain and the used representation are novel, we decided to let the evolution optimize the structure of the CNN for this particular key prediction challenge.

\section{Conclusion}\label{conclusion} In this work, we presented SnapShot; a novel oracle-less attack on logic locking based on multi-layer perceptrons and convolutional neural networks. The attack is the first of its kind that is able to immediately predict a key value out of a locked netlist using neural networks, without reconstructing the original design. Moreover, the neuroevolutionary approach showcases the agile development of a powerful attack using limited resources. We proposed a flexible representation of netlist subgraphs that can be utilized for ML-based approaches. The attack was evaluated on two realistic attack scenarios, including the previously untested generalized set scenario. SnapShot reports a significant improvement of 10.49 percentage points in key prediction accuracy, thereby achieving an average accuracy of 82.60\%; outperforming the state-of-the-art approach for all evaluated benchmarks. Furthermore, we have shown that CNNs achieve a higher performance compared to standard MLPs. Finally, we analyzed the critical vulnerabilities of logic locking that enable ML-based attacks. As a limitation, in its current format, the applicability of SnapShot to novel schemes which do not rely on XOR/XNOR gates remains to be evaluated. With SnapShot, we encourage more research on exploring the challenges of logic locking design in the deep learning era. Based on the presented insights, in future work, we plan to investigate the design of learning-resilient locking schemes. 

\bibliographystyle{ACM-Reference-Format}
\bibliography{bibliography_shortened_conf}


\begin{thebibliography}{43}


\ifx \showCODEN    \undefined \def \showCODEN     #1{\unskip}     \fi
\ifx \showDOI      \undefined \def \showDOI       #1{#1}\fi
\ifx \showISBNx    \undefined \def \showISBNx     #1{\unskip}     \fi
\ifx \showISBNxiii \undefined \def \showISBNxiii  #1{\unskip}     \fi
\ifx \showISSN     \undefined \def \showISSN      #1{\unskip}     \fi
\ifx \showLCCN     \undefined \def \showLCCN      #1{\unskip}     \fi
\ifx \shownote     \undefined \def \shownote      #1{#1}          \fi
\ifx \showarticletitle \undefined \def \showarticletitle #1{#1}   \fi
\ifx \showURL      \undefined \def \showURL       {\relax}        \fi
\providecommand\bibfield[2]{#2}
\providecommand\bibinfo[2]{#2}
\providecommand\natexlab[1]{#1}
\providecommand\showeprint[2][]{arXiv:#2}

\bibitem[\protect\citeauthoryear{{Alaql}, {Forte}, and {Bhunia}}{{Alaql}
  et~al\mbox{.}}{2019}]%
        {sweepAttack2019}
\bibfield{author}{\bibinfo{person}{A. {Alaql}}, \bibinfo{person}{D. {Forte}},
  {and} \bibinfo{person}{S. {Bhunia}}.} \bibinfo{year}{2019}\natexlab{}.
\newblock \showarticletitle{Sweep to the Secret: A Constant Propagation Attack
  on Logic Locking}. In \bibinfo{booktitle}{\emph{2019 AsianHOST}}.
  \bibinfo{pages}{1--6}.
\newblock


\bibitem[\protect\citeauthoryear{Almeida and Ludermir}{Almeida and
  Ludermir}{2008}]%
        {neuroevolution2008}
\bibfield{author}{\bibinfo{person}{Leandro~M. Almeida} {and}
  \bibinfo{person}{Teresa~B. Ludermir}.} \bibinfo{year}{2008}\natexlab{}.
\newblock \showarticletitle{An Evolutionary Approach for Tuning Artificial
  Neural Network Parameters}. In \bibinfo{booktitle}{\emph{HAIS}},
  \bibfield{editor}{\bibinfo{person}{Emilio Corchado}, \bibinfo{person}{Ajith
  Abraham}, {and} \bibinfo{person}{Witold Pedrycz}} (Eds.).
  \bibinfo{publisher}{Springer Berlin Heidelberg}, \bibinfo{address}{Berlin,
  Heidelberg}, \bibinfo{pages}{156--163}.
\newblock
\showISBNx{978-3-540-87656-4}


\bibitem[\protect\citeauthoryear{Amir, Shakya, Forte, Tehranipoor, and
  Bhunia}{Amir et~al\mbox{.}}{2017}]%
        {comparativeAnalysis}
\bibfield{author}{\bibinfo{person}{Sarah Amir}, \bibinfo{person}{Bicky Shakya},
  \bibinfo{person}{Domenic Forte}, \bibinfo{person}{Mark Tehranipoor}, {and}
  \bibinfo{person}{Swarup Bhunia}.} \bibinfo{year}{2017}\natexlab{}.
\newblock \showarticletitle{Comparative Analysis of Hardware Obfuscation for
  {IP} Protection}. In \bibinfo{booktitle}{\emph{Proceedings of GLSVLSI
  ’17}}. \bibinfo{publisher}{ACM}, \bibinfo{address}{New York, NY, USA},
  \bibinfo{pages}{363–368}.
\newblock
\showISBNx{9781450349727}
\urldef\tempurl%
\url{https://doi.org/10.1145/3060403.3060495}
\showDOI{\tempurl}


\bibitem[\protect\citeauthoryear{Baehr, Bernardini, Sigl, and
  Schlichtmann}{Baehr et~al\mbox{.}}{2019}]%
        {MLReverseEngineering}
\bibfield{author}{\bibinfo{person}{Johanna Baehr}, \bibinfo{person}{Alessandro
  Bernardini}, \bibinfo{person}{Georg Sigl}, {and} \bibinfo{person}{Ulf
  Schlichtmann}.} \bibinfo{year}{2019}\natexlab{}.
\newblock \showarticletitle{Machine Learning and Structural Characteristics for
  Reverse Engineering}. In \bibinfo{booktitle}{\emph{Proceedings of the 24th
  ASPDAC ’19}}. \bibinfo{publisher}{ACM}, \bibinfo{address}{New York, NY,
  USA}, \bibinfo{pages}{96–103}.
\newblock
\showISBNx{9781450360074}
\urldef\tempurl%
\url{https://doi.org/10.1145/3287624.3288740}
\showDOI{\tempurl}


\bibitem[\protect\citeauthoryear{Battaglia, Hamrick, Bapst, Sanchez-Gonzalez,
  Zambaldi, Malinowski, Tacchetti, Raposo, Santoro, Faulkner,
  et~al\mbox{.}}{Battaglia et~al\mbox{.}}{2018}]%
        {gnns2018}
\bibfield{author}{\bibinfo{person}{Peter~W Battaglia},
  \bibinfo{person}{Jessica~B Hamrick}, \bibinfo{person}{Victor Bapst},
  \bibinfo{person}{Alvaro Sanchez-Gonzalez}, \bibinfo{person}{Vinicius
  Zambaldi}, \bibinfo{person}{Mateusz Malinowski}, \bibinfo{person}{Andrea
  Tacchetti}, \bibinfo{person}{David Raposo}, \bibinfo{person}{Adam Santoro},
  \bibinfo{person}{Ryan Faulkner}, {et~al\mbox{.}}}
  \bibinfo{year}{2018}\natexlab{}.
\newblock \showarticletitle{Relational Inductive Biases, Deep Learning, and
  Graph Networks}.
\newblock \bibinfo{journal}{\emph{arXiv preprint arXiv:1806.01261}}
  (\bibinfo{year}{2018}).
\newblock


\bibitem[\protect\citeauthoryear{{Brglez}, {Bryan}, and {Kozminski}}{{Brglez}
  et~al\mbox{.}}{1989}]%
        {iscas2}
\bibfield{author}{\bibinfo{person}{F. {Brglez}}, \bibinfo{person}{D. {Bryan}},
  {and} \bibinfo{person}{K. {Kozminski}}.} \bibinfo{year}{1989}\natexlab{}.
\newblock \showarticletitle{Combinational Profiles of Sequential Benchmark
  Circuits}. In \bibinfo{booktitle}{\emph{IEEE ISCAS}}.
  \bibinfo{pages}{1929--1934 vol.3}.
\newblock
\urldef\tempurl%
\url{https://doi.org/10.1109/ISCAS.1989.100747}
\showDOI{\tempurl}


\bibitem[\protect\citeauthoryear{{Chakraborty}, {Cruz}, and
  {Bhunia}}{{Chakraborty} et~al\mbox{.}}{2018}]%
        {SAIL2019}
\bibfield{author}{\bibinfo{person}{P. {Chakraborty}}, \bibinfo{person}{J.
  {Cruz}}, {and} \bibinfo{person}{S. {Bhunia}}.}
  \bibinfo{year}{2018}\natexlab{}.
\newblock \showarticletitle{{SAIL}: Machine Learning Guided Structural Analysis
  Attack on Hardware Obfuscation}. In \bibinfo{booktitle}{\emph{2018
  AsianHOST}}. \bibinfo{pages}{56--61}.
\newblock
\showISSN{null}
\urldef\tempurl%
\url{https://doi.org/10.1109/AsianHOST.2018.8607163}
\showDOI{\tempurl}


\bibitem[\protect\citeauthoryear{{Chakraborty}, {Cruz}, and
  {Bhunia}}{{Chakraborty} et~al\mbox{.}}{2019}]%
        {SURF2019}
\bibfield{author}{\bibinfo{person}{P. {Chakraborty}}, \bibinfo{person}{J.
  {Cruz}}, {and} \bibinfo{person}{S. {Bhunia}}.}
  \bibinfo{year}{2019}\natexlab{}.
\newblock \showarticletitle{{SURF}: Joint Structural Functional Attack on Logic
  Locking}. In \bibinfo{booktitle}{\emph{2019 IEEE HOST}}.
  \bibinfo{pages}{181--190}.
\newblock
\showISSN{null}
\urldef\tempurl%
\url{https://doi.org/10.1109/HST.2019.8741028}
\showDOI{\tempurl}


\bibitem[\protect\citeauthoryear{{Chen}, {Fu}, {Zhao}, and {Koushanfar}}{{Chen}
  et~al\mbox{.}}{2019}]%
        {GenUnlock2019}
\bibfield{author}{\bibinfo{person}{H. {Chen}}, \bibinfo{person}{C. {Fu}},
  \bibinfo{person}{J. {Zhao}}, {and} \bibinfo{person}{F. {Koushanfar}}.}
  \bibinfo{year}{2019}\natexlab{}.
\newblock \showarticletitle{GenUnlock: An Automated Genetic Algorithm Framework
  for Unlocking Logic Encryption}. In \bibinfo{booktitle}{\emph{2019 IEEE/ACM
  ICCAD}}. \bibinfo{pages}{1--8}.
\newblock


\bibitem[\protect\citeauthoryear{{Chen}, {Kolhe}, {Rafatirad}, {Lu}, {Manoj
  P.D.}, {Homayoun}, and {Zhao}}{{Chen} et~al\mbox{.}}{2020}]%
        {EstimatingDeObfWithGNN2020}
\bibfield{author}{\bibinfo{person}{Z. {Chen}}, \bibinfo{person}{G. {Kolhe}},
  \bibinfo{person}{S. {Rafatirad}}, \bibinfo{person}{C. {Lu}},
  \bibinfo{person}{S. {Manoj P.D.}}, \bibinfo{person}{H. {Homayoun}}, {and}
  \bibinfo{person}{L. {Zhao}}.} \bibinfo{year}{2020}\natexlab{}.
\newblock \showarticletitle{Estimating the Circuit De-obfuscation Runtime based
  on Graph Deep Learning}. In \bibinfo{booktitle}{\emph{2020 DATE}}.
  \bibinfo{pages}{358--363}.
\newblock


\bibitem[\protect\citeauthoryear{{Corno}, {Reorda}, and {Squillero}}{{Corno}
  et~al\mbox{.}}{2000}]%
        {ITC99}
\bibfield{author}{\bibinfo{person}{F. {Corno}}, \bibinfo{person}{M.~S.
  {Reorda}}, {and} \bibinfo{person}{G. {Squillero}}.}
  \bibinfo{year}{2000}\natexlab{}.
\newblock \showarticletitle{{RT}-level {ITC'99} Benchmarks and First {ATPG}
  Results}.
\newblock \bibinfo{journal}{\emph{IEEE Design Test of Computers}}
  \bibinfo{volume}{17}, \bibinfo{number}{3} (\bibinfo{year}{2000}),
  \bibinfo{pages}{44--53}.
\newblock


\bibitem[\protect\citeauthoryear{Eiben and Smith}{Eiben and Smith}{2015}]%
        {Eiben2015}
\bibfield{author}{\bibinfo{person}{A.~E. Eiben} {and} \bibinfo{person}{James~E.
  Smith}.} \bibinfo{year}{2015}\natexlab{}.
\newblock \bibinfo{booktitle}{\emph{Introduction to Evolutionary Computing}
  (\bibinfo{edition}{2nd} ed.)}.
\newblock \bibinfo{publisher}{Springer Publishing Company, Incorporated}.
\newblock
\showISBNx{3662448734, 9783662448731}


\bibitem[\protect\citeauthoryear{Elsken, Metzen, and Hutter}{Elsken
  et~al\mbox{.}}{2018}]%
        {NASSurvey2018}
\bibfield{author}{\bibinfo{person}{Thomas Elsken}, \bibinfo{person}{Jan~Hendrik
  Metzen}, {and} \bibinfo{person}{Frank Hutter}.}
  \bibinfo{year}{2018}\natexlab{}.
\newblock \showarticletitle{Neural architecture search: A survey}.
\newblock \bibinfo{journal}{\emph{arXiv preprint arXiv:1808.05377}}
  (\bibinfo{year}{2018}).
\newblock


\bibitem[\protect\citeauthoryear{He, Zhao, and Chu}{He et~al\mbox{.}}{2020}]%
        {AutoMLSurvey2020}
\bibfield{author}{\bibinfo{person}{Xin He}, \bibinfo{person}{Kaiyong Zhao},
  {and} \bibinfo{person}{Xiaowen Chu}.} \bibinfo{year}{2020}\natexlab{}.
\newblock \bibinfo{title}{AutoML: A Survey of the State-of-the-Art}.
\newblock
\newblock
\showeprint[arxiv]{cs.LG/1908.00709}


\bibitem[\protect\citeauthoryear{Jain, Zhou, and Guin}{Jain
  et~al\mbox{.}}{2019}]%
        {TAAL2019}
\bibfield{author}{\bibinfo{person}{Ayush Jain}, \bibinfo{person}{Ziqi Zhou},
  {and} \bibinfo{person}{Ujjwal Guin}.} \bibinfo{year}{2019}\natexlab{}.
\newblock \bibinfo{title}{TAAL: Tampering Attack on Any Key-based Logic Locked
  Circuits}.
\newblock
\newblock
\showeprint[arxiv]{cs.CR/1909.07426}


\bibitem[\protect\citeauthoryear{{Kamali}, {Azar}, {Homayoun}, and
  {Sasan}}{{Kamali} et~al\mbox{.}}{2019}]%
        {fullLock2019}
\bibfield{author}{\bibinfo{person}{H.~M. {Kamali}}, \bibinfo{person}{K.~Z.
  {Azar}}, \bibinfo{person}{H. {Homayoun}}, {and} \bibinfo{person}{A.
  {Sasan}}.} \bibinfo{year}{2019}\natexlab{}.
\newblock \showarticletitle{Full-Lock: Hard Distributions of SAT instances for
  Obfuscating Circuits using Fully Configurable Logic and Routing Blocks}. In
  \bibinfo{booktitle}{\emph{2019 56th ACM/IEEE DAC}}. \bibinfo{pages}{1--6}.
\newblock


\bibitem[\protect\citeauthoryear{{Karmakar} and {Chattopadhyay}}{{Karmakar} and
  {Chattopadhyay}}{2020}]%
        {particleSwarmBasedLL2020}
\bibfield{author}{\bibinfo{person}{R. {Karmakar}} {and} \bibinfo{person}{S.
  {Chattopadhyay}}.} \bibinfo{year}{2020}\natexlab{}.
\newblock \showarticletitle{A Particle Swarm Optimization Guided Approximate
  Key Search Attack on Logic Locking in The Absence of Scan Access}. In
  \bibinfo{booktitle}{\emph{2020 DATE}}. \bibinfo{pages}{448--453}.
\newblock


\bibitem[\protect\citeauthoryear{Karmakar, Chattopadhyay, and Kapur}{Karmakar
  et~al\mbox{.}}{2017}]%
        {enhancing2017}
\bibfield{author}{\bibinfo{person}{R. Karmakar}, \bibinfo{person}{S.
  Chattopadhyay}, {and} \bibinfo{person}{R. Kapur}.}
  \bibinfo{year}{2017}\natexlab{}.
\newblock \showarticletitle{Enhancing Security of Logic Encryption Using
  Embedded Key Generation Unit}. In \bibinfo{booktitle}{\emph{2017 ITC-Asia}}.
  \bibinfo{pages}{131--136}.
\newblock
\urldef\tempurl%
\url{https://doi.org/10.1109/ITC-ASIA.2017.8097127}
\showDOI{\tempurl}


\bibitem[\protect\citeauthoryear{{Khan}, {Rahmani}, {Shah}, {Bennamoun},
  {Medioni}, and {Dickinson}}{{Khan} et~al\mbox{.}}{2018}]%
        {guideToCNNs2018}
\bibfield{author}{\bibinfo{person}{S. {Khan}}, \bibinfo{person}{H. {Rahmani}},
  \bibinfo{person}{S.~A.~A. {Shah}}, \bibinfo{person}{M. {Bennamoun}},
  \bibinfo{person}{G. {Medioni}}, {and} \bibinfo{person}{S. {Dickinson}}.}
  \bibinfo{year}{2018}\natexlab{}.
\newblock \bibinfo{booktitle}{\emph{A Guide to Convolutional Neural Networks
  for Computer Vision}}.
\newblock


\bibitem[\protect\citeauthoryear{LeCun and Bengio}{LeCun and Bengio}{1998}]%
        {LeCun1998}
\bibfield{author}{\bibinfo{person}{Yann LeCun} {and} \bibinfo{person}{Yoshua
  Bengio}.} \bibinfo{year}{1998}\natexlab{}.
\newblock \showarticletitle{The Handbook of Brain Theory and Neural Networks}.
\newblock \bibinfo{publisher}{MIT Press}, Chapter Convolutional Networks for
  Images, Speech, and Time Series, \bibinfo{pages}{255--258}.
\newblock
\showISBNx{0-262-51102-9}


\bibitem[\protect\citeauthoryear{Lee, Tehranipoor, and Plusquellic}{Lee
  et~al\mbox{.}}{2006}]%
        {secureScanChain}
\bibfield{author}{\bibinfo{person}{Jeremy Lee}, \bibinfo{person}{Mohammad
  Tehranipoor}, {and} \bibinfo{person}{Jim Plusquellic}.}
  \bibinfo{year}{2006}\natexlab{}.
\newblock \showarticletitle{A Low-Cost Solution for Protecting {IP}s Against
  Scan-Based Side-Channel Attacks}. In \bibinfo{booktitle}{\emph{VTS '06}}.
  \bibinfo{publisher}{IEEE Computer Society}, \bibinfo{address}{USA},
  \bibinfo{pages}{94–99}.
\newblock
\showISBNx{0769525148}
\urldef\tempurl%
\url{https://doi.org/10.1109/VTS.2006.7}
\showDOI{\tempurl}


\bibitem[\protect\citeauthoryear{{Li} and {Orailoglu}}{{Li} and
  {Orailoglu}}{2019a}]%
        {redundancyIdentification2019}
\bibfield{author}{\bibinfo{person}{L. {Li}} {and} \bibinfo{person}{A.
  {Orailoglu}}.} \bibinfo{year}{2019}\natexlab{a}.
\newblock \showarticletitle{Piercing Logic Locking Keys through Redundancy
  Identification}. In \bibinfo{booktitle}{\emph{2019 DATE}}.
  \bibinfo{pages}{540--545}.
\newblock
\showISSN{1530-1591}
\urldef\tempurl%
\url{https://doi.org/10.23919/DATE.2019.8714955}
\showDOI{\tempurl}


\bibitem[\protect\citeauthoryear{{Li} and {Orailoglu}}{{Li} and
  {Orailoglu}}{2019b}]%
        {redundancyShield2019}
\bibfield{author}{\bibinfo{person}{L. {Li}} {and} \bibinfo{person}{A.
  {Orailoglu}}.} \bibinfo{year}{2019}\natexlab{b}.
\newblock \showarticletitle{Shielding Logic Locking from Redundancy Attacks}.
  In \bibinfo{booktitle}{\emph{2019 VTS}}. \bibinfo{pages}{1--6}.
\newblock
\showISSN{1093-0167}
\urldef\tempurl%
\url{https://doi.org/10.1109/VTS.2019.8758671}
\showDOI{\tempurl}


\bibitem[\protect\citeauthoryear{Li, Yang, Peng, and Liu}{Li
  et~al\mbox{.}}{2020}]%
        {cnnSurvey2020}
\bibfield{author}{\bibinfo{person}{Zewen Li}, \bibinfo{person}{Wenjie Yang},
  \bibinfo{person}{Shouheng Peng}, {and} \bibinfo{person}{Fan Liu}.}
  \bibinfo{year}{2020}\natexlab{}.
\newblock \bibinfo{title}{A Survey of Convolutional Neural Networks: Analysis,
  Applications, and Prospects}.
\newblock
\newblock
\showeprint{arXiv:2004.02806}


\bibitem[\protect\citeauthoryear{{Liu}, {Zuzak}, {Xie}, {Chakraborty}, and
  {Srivastava}}{{Liu} et~al\mbox{.}}{2020}]%
        {strongAntiSat2020}
\bibfield{author}{\bibinfo{person}{Y. {Liu}}, \bibinfo{person}{M. {Zuzak}},
  \bibinfo{person}{Y. {Xie}}, \bibinfo{person}{A. {Chakraborty}}, {and}
  \bibinfo{person}{A. {Srivastava}}.} \bibinfo{year}{2020}\natexlab{}.
\newblock \showarticletitle{Strong Anti-SAT: Secure and Effective Logic
  Locking}. In \bibinfo{booktitle}{\emph{2020 ISQED}}.
  \bibinfo{pages}{199--205}.
\newblock


\bibitem[\protect\citeauthoryear{Massad, Zhang, Garg, and Tripunitara}{Massad
  et~al\mbox{.}}{2017}]%
        {desynthesisAttack2017}
\bibfield{author}{\bibinfo{person}{Mohamed~El Massad}, \bibinfo{person}{Jun
  Zhang}, \bibinfo{person}{Siddharth Garg}, {and} \bibinfo{person}{Mahesh~V.
  Tripunitara}.} \bibinfo{year}{2017}\natexlab{}.
\newblock \showarticletitle{Logic Locking for Secure Outsourced Chip
  Fabrication: {A} New Attack and Provably Secure Defense Mechanism}.
\newblock \bibinfo{journal}{\emph{CoRR}}  \bibinfo{volume}{abs/1703.10187}
  (\bibinfo{year}{2017}).
\newblock
\showeprint[arxiv]{1703.10187}


\bibitem[\protect\citeauthoryear{Rahman, Tajik, Rahman, Tehranipoor, and
  Asadizanjani}{Rahman et~al\mbox{.}}{2019}]%
        {KeyReading2019}
\bibfield{author}{\bibinfo{person}{Mir~Tanjidur Rahman},
  \bibinfo{person}{Shahin Tajik}, \bibinfo{person}{M.~Sazadur Rahman},
  \bibinfo{person}{Mark Tehranipoor}, {and} \bibinfo{person}{Navid
  Asadizanjani}.} \bibinfo{year}{2019}\natexlab{}.
\newblock \bibinfo{title}{The Key is Left under the Mat: On the Inappropriate
  Security Assumption of Logic Locking Schemes}.
\newblock \bibinfo{howpublished}{Cryptology ePrint Archive, Report 2019/719}.
\newblock
\newblock
\shownote{\url{https://eprint.iacr.org/2019/719}.}


\bibitem[\protect\citeauthoryear{{Rajendran}, {Pino}, {Sinanoglu}, and
  {Karri}}{{Rajendran} et~al\mbox{.}}{2012}]%
        {sll1}
\bibfield{author}{\bibinfo{person}{J. {Rajendran}}, \bibinfo{person}{Y.
  {Pino}}, \bibinfo{person}{O. {Sinanoglu}}, {and} \bibinfo{person}{R.
  {Karri}}.} \bibinfo{year}{2012}\natexlab{}.
\newblock \showarticletitle{Security Analysis of Logic Obfuscation}. In
  \bibinfo{booktitle}{\emph{DAC 2012}}. \bibinfo{pages}{83--89}.
\newblock
\showISSN{0738-100X}
\urldef\tempurl%
\url{https://doi.org/10.1145/2228360.2228377}
\showDOI{\tempurl}


\bibitem[\protect\citeauthoryear{Rostami, Koushanfar, and Karri}{Rostami
  et~al\mbox{.}}{2014}]%
        {rostami2014primer}
\bibfield{author}{\bibinfo{person}{M. Rostami}, \bibinfo{person}{F.
  Koushanfar}, {and} \bibinfo{person}{R. Karri}.}
  \bibinfo{year}{2014}\natexlab{}.
\newblock \showarticletitle{A Primer on Hardware Security: Models, Methods, and
  Metrics}.
\newblock \bibinfo{journal}{\emph{Proc. IEEE}} \bibinfo{volume}{102},
  \bibinfo{number}{8} (\bibinfo{date}{Aug} \bibinfo{year}{2014}),
  \bibinfo{pages}{1283--1295}.
\newblock
\showISSN{0018-9219}


\bibitem[\protect\citeauthoryear{{Roy}, {Koushanfar}, and {Markov}}{{Roy}
  et~al\mbox{.}}{2008}]%
        {epic2010}
\bibfield{author}{\bibinfo{person}{J.~A. {Roy}}, \bibinfo{person}{F.
  {Koushanfar}}, {and} \bibinfo{person}{I.~L. {Markov}}.}
  \bibinfo{year}{2008}\natexlab{}.
\newblock \showarticletitle{{EPIC}: Ending Piracy of Integrated Circuits}. In
  \bibinfo{booktitle}{\emph{2008 DATE}}. \bibinfo{pages}{1069--1074}.
\newblock
\showISSN{1558-1101}
\urldef\tempurl%
\url{https://doi.org/10.1109/DATE.2008.4484823}
\showDOI{\tempurl}


\bibitem[\protect\citeauthoryear{Shakya, Xu, Tehranipoor, and Forte}{Shakya
  et~al\mbox{.}}{2019}]%
        {caslock2019}
\bibfield{author}{\bibinfo{person}{Bicky Shakya}, \bibinfo{person}{Xiaolin Xu},
  \bibinfo{person}{Mark Tehranipoor}, {and} \bibinfo{person}{Domenic Forte}.}
  \bibinfo{year}{2019}\natexlab{}.
\newblock \showarticletitle{CAS-Lock: A Security-Corruptibility Trade-off
  Resilient Logic Locking Scheme}.
\newblock \bibinfo{journal}{\emph{TCHES}} \bibinfo{volume}{2020},
  \bibinfo{number}{1} (\bibinfo{date}{Nov.} \bibinfo{year}{2019}),
  \bibinfo{pages}{175--202}.
\newblock
\urldef\tempurl%
\url{https://doi.org/10.13154/tches.v2020.i1.175-202}
\showDOI{\tempurl}


\bibitem[\protect\citeauthoryear{Simonyan and Zisserman}{Simonyan and
  Zisserman}{2015}]%
        {cnn2015}
\bibfield{author}{\bibinfo{person}{Karen Simonyan} {and}
  \bibinfo{person}{Andrew Zisserman}.} \bibinfo{year}{2015}\natexlab{}.
\newblock \showarticletitle{Very Deep Convolutional Networks for Large-Scale
  Image Recognition}. In \bibinfo{booktitle}{\emph{ICLR 2015, San Diego, CA,
  USA, May 7-9, 2015, Conference Track Proceedings}},
  \bibfield{editor}{\bibinfo{person}{Yoshua Bengio} {and} \bibinfo{person}{Yann
  LeCun}} (Eds.).
\newblock
\urldef\tempurl%
\url{http://arxiv.org/abs/1409.1556}
\showURL{%
\tempurl}


\bibitem[\protect\citeauthoryear{{Subramanyan}, {Ray}, and
  {Malik}}{{Subramanyan} et~al\mbox{.}}{2015}]%
        {sat2015}
\bibfield{author}{\bibinfo{person}{P. {Subramanyan}}, \bibinfo{person}{S.
  {Ray}}, {and} \bibinfo{person}{S. {Malik}}.} \bibinfo{year}{2015}\natexlab{}.
\newblock \showarticletitle{Evaluating the Security of Logic Encryption
  Algorithms}. In \bibinfo{booktitle}{\emph{2015 HOST}}.
  \bibinfo{pages}{137--143}.
\newblock
\showISSN{null}
\urldef\tempurl%
\url{https://doi.org/10.1109/HST.2015.7140252}
\showDOI{\tempurl}


\bibitem[\protect\citeauthoryear{Szegedy, Ioffe, Vanhoucke, and Alemi}{Szegedy
  et~al\mbox{.}}{2017}]%
        {cnn2017}
\bibfield{author}{\bibinfo{person}{Christian Szegedy}, \bibinfo{person}{Sergey
  Ioffe}, \bibinfo{person}{Vincent Vanhoucke}, {and}
  \bibinfo{person}{Alexander~A. Alemi}.} \bibinfo{year}{2017}\natexlab{}.
\newblock \showarticletitle{Inception-v4, Inception-ResNet and the Impact of
  Residual Connections on Learning}. In \bibinfo{booktitle}{\emph{AAAI 2017}}.
  \bibinfo{publisher}{AAAI Press}, \bibinfo{pages}{4278–4284}.
\newblock


\bibitem[\protect\citeauthoryear{Tehranipoor, Karimian, Mozaffari~Kermani, and
  Mahmoodi}{Tehranipoor et~al\mbox{.}}{2019}]%
        {BOCANet2019}
\bibfield{author}{\bibinfo{person}{Fatemeh Tehranipoor}, \bibinfo{person}{Nima
  Karimian}, \bibinfo{person}{Mehran Mozaffari~Kermani}, {and}
  \bibinfo{person}{Hamid Mahmoodi}.} \bibinfo{year}{2019}\natexlab{}.
\newblock \showarticletitle{Deep {RNN}-Oriented Paradigm Shift through
  {BOCANet}: Broken Obfuscated Circuit Attack}. In
  \bibinfo{booktitle}{\emph{GLSVLSI ’19}}. \bibinfo{publisher}{ACM},
  \bibinfo{address}{New York, NY, USA}, \bibinfo{pages}{335–338}.
\newblock
\showISBNx{9781450362528}
\urldef\tempurl%
\url{https://doi.org/10.1145/3299874.3318031}
\showDOI{\tempurl}


\bibitem[\protect\citeauthoryear{\v{S}i\v{s}ejkovi\'{c}, Merchant, Reimann,
  Leupers, Giacometti, and Kegrei\ss{}}{\v{S}i\v{s}ejkovi\'{c}
  et~al\mbox{.}}{2020}]%
        {sisejkovic2020scopes}
\bibfield{author}{\bibinfo{person}{Dominik \v{S}i\v{s}ejkovi\'{c}},
  \bibinfo{person}{Farhad Merchant}, \bibinfo{person}{Lennart~M. Reimann},
  \bibinfo{person}{Rainer Leupers}, \bibinfo{person}{Massimiliano Giacometti},
  {and} \bibinfo{person}{Sascha Kegrei\ss{}}.} \bibinfo{year}{2020}\natexlab{}.
\newblock \showarticletitle{A Secure Hardware-Software Solution Based on
  RISC-V, Logic Locking and Microkernel}. In \bibinfo{booktitle}{\emph{SCOPES
  ’20}}. \bibinfo{publisher}{ACM}, \bibinfo{address}{New York, NY, USA},
  \bibinfo{pages}{62–65}.
\newblock
\showISBNx{9781450371315}
\urldef\tempurl%
\url{https://doi.org/10.1145/3378678.3391886}
\showDOI{\tempurl}


\bibitem[\protect\citeauthoryear{Xiao, Forte, Jin, Karri, Bhunia, and
  Tehranipoor}{Xiao et~al\mbox{.}}{2016}]%
        {trojansLessons}
\bibfield{author}{\bibinfo{person}{K. Xiao}, \bibinfo{person}{D. Forte},
  \bibinfo{person}{Y. Jin}, \bibinfo{person}{R. Karri}, \bibinfo{person}{S.
  Bhunia}, {and} \bibinfo{person}{M. Tehranipoor}.}
  \bibinfo{year}{2016}\natexlab{}.
\newblock \showarticletitle{Hardware Trojans: Lessons Learned after One Decade
  of Research}.
\newblock \bibinfo{journal}{\emph{ACM Trans. Des. Autom. Electron. Syst.}}
  \bibinfo{volume}{22}, \bibinfo{number}{1}, Article \bibinfo{articleno}{6}
  (\bibinfo{date}{May} \bibinfo{year}{2016}), \bibinfo{numpages}{23}~pages.
\newblock
\showISSN{1084-4309}
\urldef\tempurl%
\url{https://doi.org/10.1145/2906147}
\showDOI{\tempurl}


\bibitem[\protect\citeauthoryear{Xie and Srivastava}{Xie and
  Srivastava}{2017}]%
        {antisat2017}
\bibfield{author}{\bibinfo{person}{Y. Xie} {and} \bibinfo{person}{A.
  Srivastava}.} \bibinfo{year}{2017}\natexlab{}.
\newblock \bibinfo{title}{Anti-SAT: Mitigating SAT Attack on Logic Locking}.
\newblock \bibinfo{howpublished}{Cryptology ePrint Archive, Report 2017/761}.
\newblock
\newblock
\shownote{\url{http://eprint.iacr.org/2017/761}.}


\bibitem[\protect\citeauthoryear{{Yasin} and {Sinanoglu}}{{Yasin} and
  {Sinanoglu}}{2017}]%
        {evo2017}
\bibfield{author}{\bibinfo{person}{M. {Yasin}} {and} \bibinfo{person}{O.
  {Sinanoglu}}.} \bibinfo{year}{2017}\natexlab{}.
\newblock \showarticletitle{Evolution of Logic Locking}. In
  \bibinfo{booktitle}{\emph{2017 IFIP/IEEE VLSI-SoC}}. \bibinfo{pages}{1--6}.
\newblock
\showISSN{2324-8440}
\urldef\tempurl%
\url{https://doi.org/10.1109/VLSI-SoC.2017.8203496}
\showDOI{\tempurl}


\bibitem[\protect\citeauthoryear{Zamiri~Azar, Mardani~Kamali, Homayoun, and
  Sasan}{Zamiri~Azar et~al\mbox{.}}{2019}]%
        {decadeOfLocking}
\bibfield{author}{\bibinfo{person}{Kimia Zamiri~Azar}, \bibinfo{person}{Hadi
  Mardani~Kamali}, \bibinfo{person}{Houman Homayoun}, {and}
  \bibinfo{person}{Avesta Sasan}.} \bibinfo{year}{2019}\natexlab{}.
\newblock \showarticletitle{Threats on Logic Locking: A Decade Later}. In
  \bibinfo{booktitle}{\emph{GLSVLSI ’19}}. \bibinfo{publisher}{ACM},
  \bibinfo{address}{New York, NY, USA}, \bibinfo{pages}{471–476}.
\newblock
\showISBNx{9781450362528}
\urldef\tempurl%
\url{https://doi.org/10.1145/3299874.3319495}
\showDOI{\tempurl}


\bibitem[\protect\citeauthoryear{{Zaruba} and {Benini}}{{Zaruba} and
  {Benini}}{2019}]%
        {ariane2019}
\bibfield{author}{\bibinfo{person}{F. {Zaruba}} {and} \bibinfo{person}{L.
  {Benini}}.} \bibinfo{year}{2019}\natexlab{}.
\newblock \showarticletitle{The Cost of Application-Class Processing: Energy
  and Performance Analysis of a {L}inux-Ready 1.7-{GH}z 64-Bit {RISC-V} Core in
  22-nm {FDSOI} Technology}.
\newblock \bibinfo{journal}{\emph{IEEE Transactions on Very Large Scale
  Integration (VLSI) Systems}} \bibinfo{volume}{27}, \bibinfo{number}{11}
  (\bibinfo{date}{Nov} \bibinfo{year}{2019}), \bibinfo{pages}{2629--2640}.
\newblock
\showISSN{1557-9999}
\urldef\tempurl%
\url{https://doi.org/10.1109/TVLSI.2019.2926114}
\showDOI{\tempurl}


\bibitem[\protect\citeauthoryear{Zhang, He, and Katabi}{Zhang
  et~al\mbox{.}}{2019b}]%
        {GNNDistributedCircuitDesign2019}
\bibfield{author}{\bibinfo{person}{Guo Zhang}, \bibinfo{person}{Hao He}, {and}
  \bibinfo{person}{Dina Katabi}.} \bibinfo{year}{2019}\natexlab{b}.
\newblock \showarticletitle{Circuit-{GNN}: Graph Neural Networks for
  Distributed Circuit Design} \emph{(\bibinfo{series}{Proceedings of Machine
  Learning Research})}, \bibfield{editor}{\bibinfo{person}{Kamalika Chaudhuri}
  {and} \bibinfo{person}{Ruslan Salakhutdinov}} (Eds.),
  Vol.~\bibinfo{volume}{97}. \bibinfo{publisher}{PMLR}, \bibinfo{address}{Long
  Beach, California, USA}, \bibinfo{pages}{7364--7373}.
\newblock
\urldef\tempurl%
\url{http://proceedings.mlr.press/v97/zhang19e.html}
\showURL{%
\tempurl}


\bibitem[\protect\citeauthoryear{Zhang, Cui, Zhou, and Guin}{Zhang
  et~al\mbox{.}}{2019a}]%
        {TGA2019}
\bibfield{author}{\bibinfo{person}{Yuqiao Zhang}, \bibinfo{person}{Pinchen
  Cui}, \bibinfo{person}{Ziqi Zhou}, {and} \bibinfo{person}{Ujjwal Guin}.}
  \bibinfo{year}{2019}\natexlab{a}.
\newblock \showarticletitle{TGA: An Oracle-Less and Topology-Guided Attack on
  Logic Locking}. In \bibinfo{booktitle}{\emph{ASHES 2019}}.
  \bibinfo{publisher}{ACM}, \bibinfo{address}{New York, NY, USA},
  \bibinfo{pages}{75–83}.
\newblock
\showISBNx{9781450368391}
\urldef\tempurl%
\url{https://doi.org/10.1145/3338508.3359576}
\showDOI{\tempurl}


\end{thebibliography}

\end{document}